\documentclass[letterpaper,11pt]{article}

\usepackage{jhepmod}

\def\be{\begin{equation}}
\def\ee{\end{equation}}
\def\ben{\begin{eqnarray}}
\def\een{\end{eqnarray}}
\def\nn{\nonumber}
\def\oh{\bf \hat\Omega}

\def\bk{{\bf k}}

\def\bk{{\bf k}}

\def\bl{{\bf l}}

\def\2p{{(2\pi)^2}}

\def\bl{{\bf l}}
\def\be{\begin{equation}}
\def\ee{\end{equation}}
\def\beq{\begin{equation}}
\def\eeq{\end{equation}}
\def\ben{\begin{eqnarray}}
\def\een{\end{eqnarray}}
\def\bes{\begin{subequations}}
\def\ees{\end{subequations}}

\def\oh{{\hat\Omega}}
\def\nn{{\nonumber}}

\def\ikap0{{\cal J}_{\theta_0}(r)}

\def\one1{\langle \kappa_{(i)}\kappa_{(j)} \rangle}
\def\one{{[\bar \xi^{(ij)}]}}

\def\ba{\begin{eqnarray}}
\def\ea{\end{eqnarray}}

\def\bk{{\bf k}}

\def\bk{{\bf k}}

\def\bl{{\bf l}}

\def\2p{{(2\pi)^2}}

\def\bl{{\bf l}}

\def\be{\begin{equation}}
\def\ee{\end{equation}}

\def\beq{\begin{equation}}
\def\eeq{\end{equation}}

\def\ben{\begin{eqnarray}}
\def\een{\end{eqnarray}}

\def\oh{{\hat\Omega}}
\def\nn{{\nonumber}}

\def\bk{{\bf k}}

\def\bl{{\bf l}}

\def\2p{{(2\pi)^2}}

\def\bl{{\bf l}}

\def\bl{{\bf{l}}}

\def\th{{\boldsymbol\theta}}

\usepackage{booktabs}
\usepackage[english]{babel}
\usepackage{amsmath,amssymb,amsbsy,amstext, amsthm, simplewick, amsfonts}
\usepackage{hyperref}
\usepackage{graphicx}
\usepackage{subcaption}
\usepackage{siunitx}
\usepackage{upgreek}
\usepackage{framed}
\usepackage{wrapfig}
\usepackage{multirow}
\usepackage{bbm}
\usepackage[svgnames,dvipsnames,x11names]{xcolor}
\usepackage[utf8x]{inputenc}
\usepackage{selinput}
\usepackage{bm}
\usepackage{float}
\usepackage{yfonts}
\usepackage{mathtools}
\usepackage{enumitem}   
\usepackage{longtable}
\usepackage{tabularx}

\usepackage{comment}

\usepackage{xcolor}
\usepackage{soul}

\usepackage{xcolor}

\def\nn{\nonumber\\}

\def\beq{\begin{equation}}
\def\eeq{\end{equation}}
\def\be{\begin{equation}}
\def\ee{\end{equation}}

\def\deltaf{
  \left (\sum^{\infty}_{n=0} {1 \over n!} f_n [\delta_{L}]^n \right )
}

\def\th{{\boldsymbol\theta}}

\def\delL{[{\hat\delta_L}]^n}

\newcommand{\rtrv}[1]{{\textcolor{black}{#1}}} % RT comment
\newcommand{\alan}[1]{{\textcolor{Blue}{#1}}} % AH comment

\title{On Weak Lensing Response Functions}
\author{D. Munshi$^{a,b,1}$, R. Takahashi$^{c,2}$, J. D. McEwen$^{b,3}$}
\affiliation{$^{a}$Imperial Centre for Inference and Cosmology (ICIC) \& Astrophysics group, \\
  Imperial College, Blackett Laboratory, Prince Consort Road, London SW7 2AZ, UK \\
$^{b}$Mullard Space Science Laboratory, University College London, \\
  Holmbury St Mary, Dorking, Surrey RH5 6NT, UK \\
  $^{c}$Faculty of Science and Technology, Hirosaki University, \\
  3 Bunkyo-cho, Hirosaki, Aomori, 036-8561, Japan \\}
\emailAdd{$^1$D.Munshi@ucl.ac.uk,
  $^{b}$takahasi@hirosaki-u.ac.jp
  $^3$Jason.McEwen@ucl.ac.uk}
\abstract{We introduce the response function (RFs) approach 
  to model the weak lensing statistics in the context of
  separate universe formalism.
  Numerical results for the RFs are presented for various
  semi-analytical models
  that includes perturbative modelling and variants of halo models.
  These results extend the recent
  studies of the Integrated Bispectrum (IB) and Trispectrum to
  arbitrary order. We find that
  due to the line-of-sight (los) projection effects, the expressions for RFs are
  not identical to the squeezed correlation functions of the same order.
  We compute the RFs in three-dimensions (3D) using the spherical
  Fourier-Bessel (sFB) formalism
  which provides a natural framework for
  incorporating photometric redshifts, and relate these expressions to tomographic and projected statistics.
  We generalise the concept of $k$-cut power spectrum to $k$-cut response functions.
  In addition to response function for high-order spectra, we also
  define their counterparts in real space, since they are easier to
  estimate from surveys with low sky-coverage and non-trivial survey boundaries.
 }

\begin{document}
%%%%%%%%%%%%%%%%%
\maketitle
%
%%%%%%%%%%%%%%%%%%%%%
\section{Introduction}
\label{sec:intro}
%%%%%%%%%%%%%%%%%%%%%%

The current generation of weak lensing surveys \citep{Review1, MunshiReview} including the
{Subaru Hypersuprimecam survey}\footnote{\href{http://www.naoj.org/Projects/HSC/index.html}{\tt http://www.naoj.org/Projects/HSC/index.html}}(HSC)
\citep{HSC},
Dark Energy Survey\footnote{\href{https://www.darkenergysurvey.org/}{\tt https://www.darkenergysurvey.org/}}(DES)\citep{DES},
Dark Energy Spectroscopic Instruments (DESI)\footnote{\href{http://desi.lbl.gov}{\tt http://desi.lbl.gov}},
Prime Focus Spectrograph\footnote{\href{http://pfs.ipmu.jp}{\tt http://pfs.ipmu.jp}},
KiDS\citep{KIDS} are already able to put
cosmological constraints that are competitive with recent
Cosmic Microwave Background surveys.
The near-future Stage-IV large scale structure (LSS)
surveys such as \textit{Euclid}\footnote{\href{http://sci.esa.int/euclid/}{\tt http://sci.esa.int/euclid/}}\citep{Euclid},
Rubin Observatory\footnote{\href{http://www.lsst.org/lsst home.shtml}{\tt {http://www.lsst.org/lsst home.shtml}}}\citep{LSSTTyson} and Roman Space Telescope\citep{WFIRST}
will improve the constraints by an order-of-magnitude and
provide answers to many of the questions
that cosmology is facing. These will provide answers to many outstanding cosmological questions, including but not limited to, nature of dark matter (DM),
dark energy (DE), possible modifications of General Relativity (GR) on cosmological scales \citep{MG1,MG2} and the sum of the neutrino masses \citep{nu}. 

Weak lensing observations target the relatively low-redshift ($z\sim 1$)
universe and small scales where the perturbations are in the nonlinear regime and
their statistics are
non-Gaussian \citep{Review1, bernardeaureview, MunshiReview}.
Indeed, understanding higher-order statistics is important as they
can significantly reduce the degeneracy
in cosmological parameters\citep{Lombardi}.
Nevertheless, higher-order statistics beyond the bispectrum and trispectrum are known to be difficult to model
analytically, and in perturbation theory higher-order contributions becomes increasingly intractable as the order increases.
Analytical modelling of the
weak lensing three-point correlation function
was initiated in real-space in  \citep{Lombardi, schneider_Kilbinger_Lombard}, and parallel
development in the harmonic domain was initiated in \citep{TakadaJain,KayoTakada}.
For early detection of non-Gaussianity see \citep{Semoloni}. 

Another strand of work has involved designing and optimising
estimators of non-Gaussianity.
The numerical estimators are computationally demanding to implement.
In addition, higher-order estimators are typically noise-dominated on small scales
and cosmic variance dominated on large scales. A large number of
simulations are required to accurate characterization \citep{Sabino_review}.
Many different estimators have recently been proposed which probe the
higher-order statistics of weak lensing maps \citep{Carbone}. These include
the well-known real-space one-point statistics such as the
cumulants \citep{Barber1} or their two-point correlators also known as
the cumulant correlators as well as the associated PDF \citep{Uhlemann1} and
the peak-count statistics \citep{peak_count}.
In the harmonic domain 
the estimators such as the Skew-Spectrum\citep{skew},
Integrated Bispectrum \citep{Integrated}
kurt-spectra \citep{kurt}, morphological estimator \citep{morph},
integrated trispectrum \citep{IT},
Betti number \citep{Betti},
extreme value statistics \citep{EVS}, position-dependent PDF \citep{pPDF},
density split statistics \citep{split}, response function formalism \citep{response}, statistics of phase \citep{Matsubara1,Matsubara2, MyPhase},
estimators for shapes of the lensing bispectrum \citep{shape}
are some of the statistical estimators
and formalism recently considered by various authors in the context of understanding
cosmological statistics in general and weak lensing in particular.
In recent years approaches based on machine learning have
also been employed \citep{MachineLearning}.

In recent years a novel technique known as a Separate Universe (SU) formalism
was developed by many authors (see e.g. \citep{Thesis} for a complete list of
references). The primary aim of this paper is to introduce the SU formalism in the
study of a specific weak lensing statistic known as the response function.
We will use this statistic to probe non-Gaussianity in weak lensing maps in projection (2D) as well as in three dimensions (3D).
We will show due to projection effects, the response functions are not
identical to the higher-order correlation functions in the squeezed limit but are closely related.

This paper is arranged as follows. In \textsection\ref{sec:intro} we introduce
our notations, next, in \ref{sec:weak} we detail the
formalism of response functions in the context of separate Universe formalism
for weak lensing convergence.
In \textsection\ref{sec:3D} we develop the response functions
for the weak lensing. The response functions for higher-order
cross-correlations against CMB is given in \textsection\ref{sec:cross}. 
Response functions for $k$-cut correlation functions are presented
in \textsection\ref{sec:kcut}. The results are discussed in \textsection\ref{sec:results} and conclusions and future prospects
are presented in \textsection\ref{sec:conclu}.

The cosmological model parameters used are the
Planck2015 best-fit flat $\Lambda$CDM model \cite{Planck2015}:
$h=0.6727$, $\Omega_b=0.0492$, $\Omega_m=0.3156$,
$\Omega_\Lambda=0.6844$, $n_s=0.9645$ and $\sigma_8=0.831$.

%/Users/mun/ONGOING/Separate_Universe/data_Q
\begin{figure}
  \begin{center}
  \begin{minipage}[b]{0.3\textwidth}
    \includegraphics[width=\textwidth]{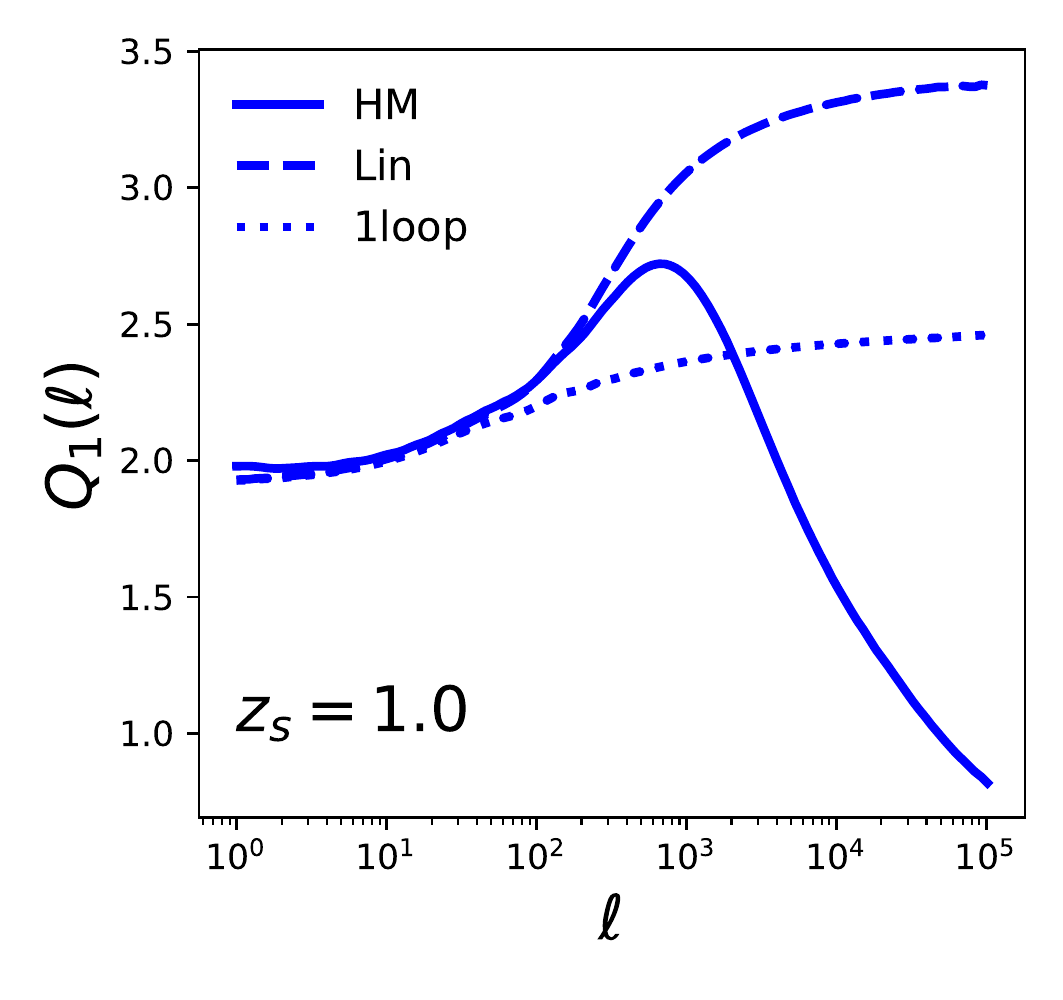}
    %\label{fig:1}
  \end{minipage}
  \begin{minipage}[b]{0.3\textwidth}
    \includegraphics[width=\textwidth]{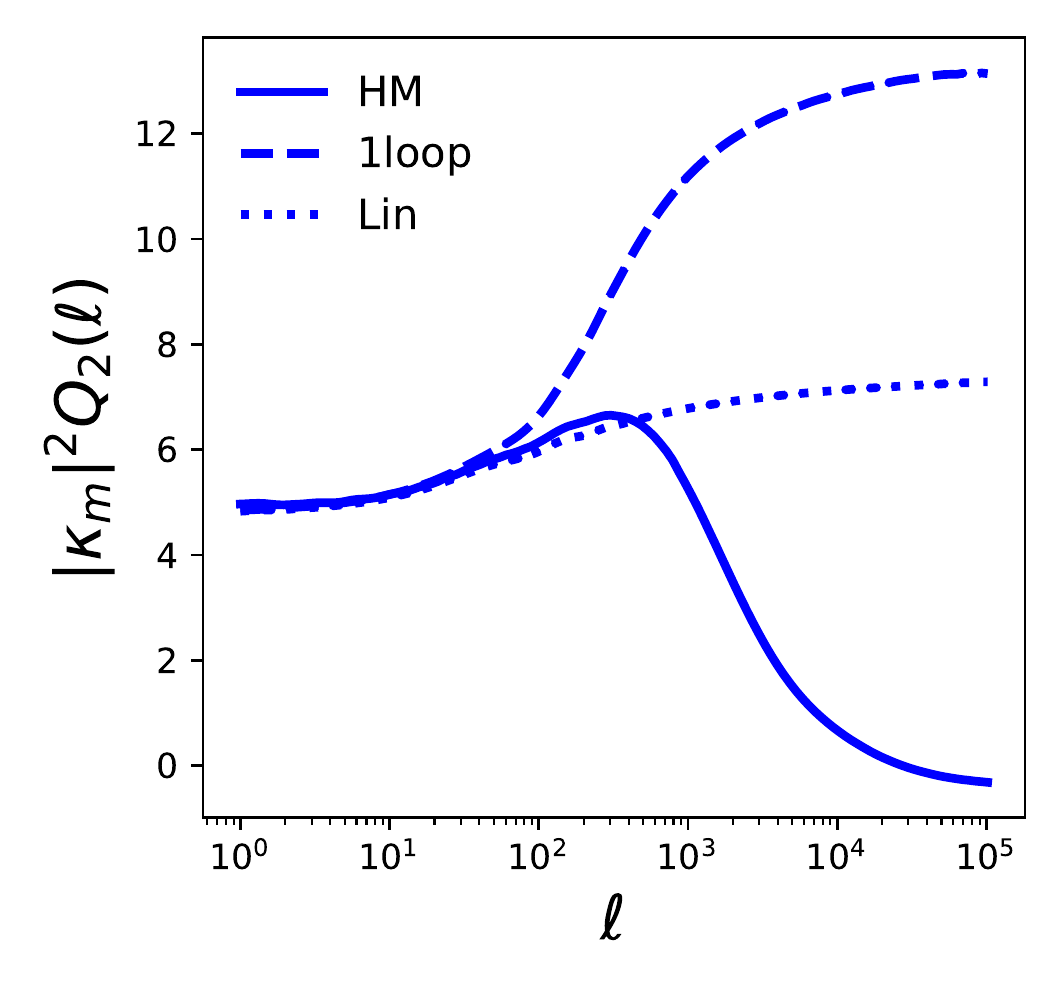}
    %\label{fig:2}
  \end{minipage}
  \begin{minipage}[b]{0.3\textwidth}
    \includegraphics[width=\textwidth]{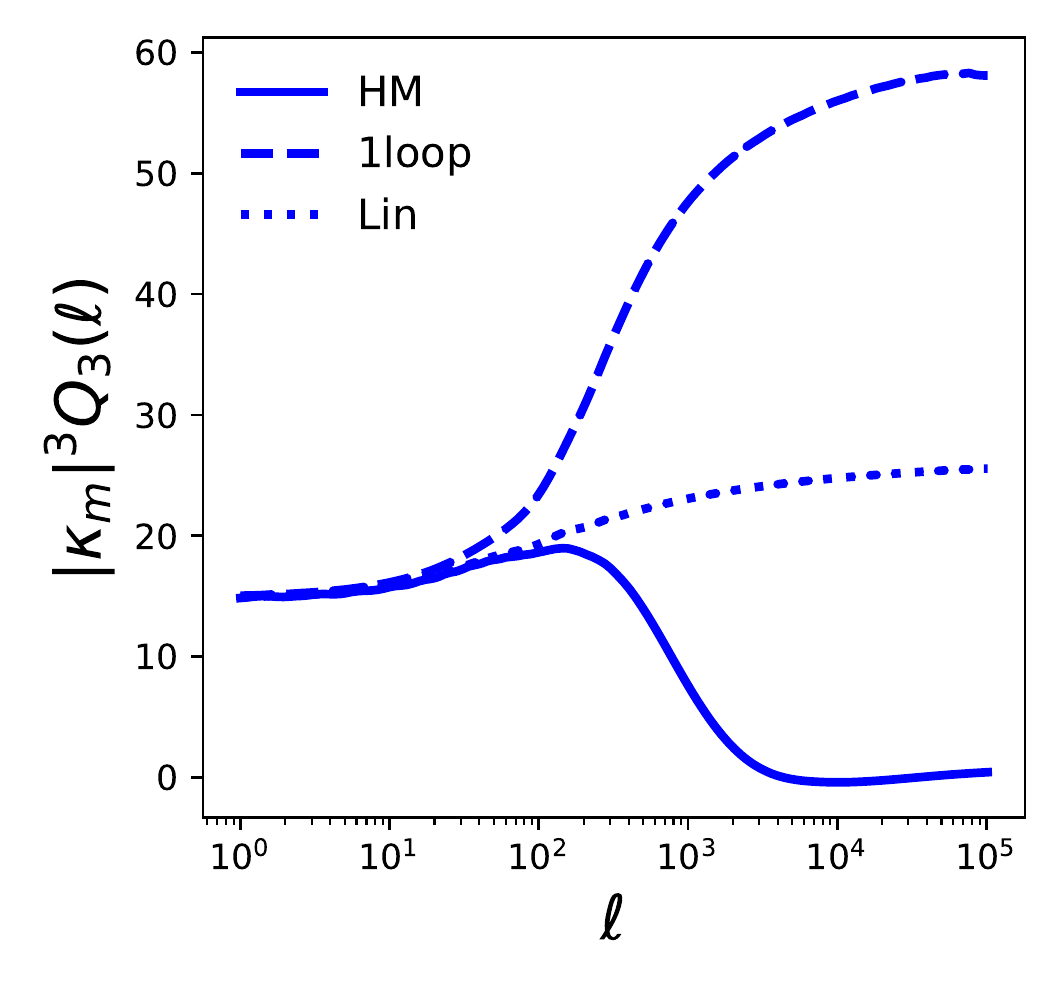}
    %\label{fig:2}
  \end{minipage}
  \caption{The response functions $|\kappa_m|^NQ_N$ defined in Eq.(\ref{eq:def_QN}) are shown. From
    left to right panels depict $N=1,2$ and $N=3$. The source redshift is $z_s=1$.
    Various line styles correspond to different analytical
    models, linear, one-loop and halo model as indicated (see text for details).
    }
  \label{fig:QN_z1}
  \end{center}
\end{figure}
%
%%%%%%%%%%%%%%%%%%%%%%%%%%%%%%%%%%%%%%%%%%%%%%%%%%%%%%%%
\section{Weak Lensing Convergence in Projection}
\label{sec:weak}
%%%%%%%%%%%%%%%%%%%%%%%%%%%%%%%%%%%%%%%%%%%%%%%%%%%%%%%%
The projected (2D) weak lensing convergence $\kappa$ is a line-of-sight integration of the
underlying three-dimensional (3D) cosmological density contrast $\delta$. The $\kappa({\th})$
at a position $\th$ can be expressed as follows: 
\bes
\ben
&& \kappa({\bm\theta}, z_s) := \int_0^{r_s} dr W(r) \delta(r,{\bm\theta}); \label{eq:singe_source}\\
&& := {3 \Omega_{\rm M} \over 2} {H_0^2 \over c^2} a^{-1} {d_A(r) d_{A}({r_s-r}) \over d_A(r_s)}.
\label{eq:w}
\een
\ees
We will suppress the variables $r_s$
unless we consider the case where
the sources are not confined 
in a single source plane.
Here $d_A(r)$ is the comoving angular diameter distance at a comoving distance $r$, i.e., $\kappa({\bm\theta}, z_s) \equiv 
\kappa({\bm\theta})$ and 
$W(r.r_s) \equiv W(r)$. 
The kernel $W(r)$ encodes geometrical dependence;
$a$ is the scale factor, $H_0$ is the Hubble constant and $\Omega_M$ is the
cosmological density parameter. $d_A(r)$ and $d_A(r_s)$
are comoving angular diameter distances at a comoving
distances $r$ and $r_s$.  We have assumed all sources to be at a single
source plane at a distance $r_s$. The projected lensing power spectrum ${\cal C}_\ell$ is given in the Limber and Born approximations by:
\ben
&& {\cal C}_{\ell} = \int_0^{r_s} dr {W^2(r) \over d_A^2(r)} 
P_{\delta}\left ({{\ell} \over d_A(r)}; r \right ).
\label{eq:global_ps}
\een
The tomographic power spectrum ${\cal C}^{ij}_{\ell}$ is given by restricting the line-of-sight
integration for sources in a particular estimated redshift bins (labelled by $i,j$):
\ben
&& {\cal C}^{ij}_{\ell} = \int_0^{r_s} dr {W_i(r) 
  W_j(r) \over d_A^2(r)} P_{\delta}\left ({{\ell} \over d_A(r)}; r \right ).
\label{eq:cross_spectra_define}
\een
the kernels $W_i$ and $W_j$ can be obtained by replacing $r_s$ in Eq.(\ref{eq:w}) respectively by $r_{si}$
and $r_j$. The upperlimit
of integration will be
$r_s = min(r_{si}, r_{sj})$.

Throughout, we assume that all sources are at a single source
redshift distribution. For a distribution of source redshifts
there is a further radial integration.
\ben
&& {\cal C}_{\ell} = \int_0^{r_s} dr {{\hat W}^2(r) \over d_A^2(r)} 
P_{\delta}\left ({{\ell} \over d_A(r)}; r \right ).
\label{eq:global_ps}
\een
\ben
&& \kappa(\bm\theta) = 
\int_0^{\infty}
n(z_s){\kappa(\bm\theta, z_s)dz_s}; \quad
=\int_0^{r_{max}}
dr \hat W(r)\delta(r); \\
&& \hat W(r) = 
\int^{\infty}_{z_s} dz_s W(r,r_s) n(z_s).
\label{eq:define_hat}
\een
The integral along the radial direction takes 
into account contribution from 
individual source planes and $n(z_s)$
number density of sources at a source
redshift $z_s$. We ignore the discreteness effect.

%
%%%%%%%%%%%%%%%%%%%%%%%%%%%%%%%%%%%%%%%%%%%%%%%%%%%%%%%%%%%%%%%%%%%%%%%%%
\section{Separate Universe Formalism and Response Functions in Projection}
%%%%%%%%%%%%%%%%%%%%%%%%%%%%%%%%%%%%%%%%%%%%%%%%%%%%%%%%%%%%%%%%%%%%%%%%%
In a SU formalism an infinite wavelength
adiabatic perturbation $\delta_{\rho}$
is absorbed in the background matter density by redefining
the cosmological parameters \rtrv{(e.g., \cite{baldauf2011,TH2013})}. We will denote the
comoving coordinate, scale-factor, comoving wave number,
and power spectrum respectively as ${\bf x}$, $a(t)$, $k$ and $P(k)$.
The corresponding quantities in the modified cosmology will be denoted as
$\tilde{\bf x}$, $\tilde a(t)$, $\tilde k$ and $\tilde P(\tilde k)$.
We will also introduce $\delta_a$ and $\delta_\rho$ the
 Lagrangian and Eulerian perturbation as follows:
\bes
\ben
&&{\bf x} a(t) = \tilde{\bf x}{\tilde a}(t); \quad
1+ \delta_a = {\tilde a(t)\over a(t)}\\
&& (1+\delta_\rho) = (1+\delta_a)^{-3}. \\
&& \tilde\Omega_M \tilde h^2 = \Omega_M h^2 \\
&& \tilde x = (1+\delta_a)^{-1}x; \quad\quad \tilde k = (1+\delta_a)k.
\een
\ees
Following the derivation in \citep{Thesis} we can expand $\delta_a$ and $\delta_\rho$ in terms of the linear overdensity $\delta_L = D_+\delta_{L0}$:
The evolution of $\delta_a$ and $\delta_\rho$
in the fiducial
cosmology can be solved using a spherical collapse model. 
The equations are further simplified by an Einstein de Sittter (EdS)
cosmology. The accuracy of such approximations have been tested 
and was found to be better than a few percents.

\bes
\ben
&& \delta_a = \sum_{n=1}^{\infty} e_n [\delta_L]^n; \quad
\delta_\rho = \sum_{n=1}^{\infty} f_n [\delta_L]^n; \\
&& e_i = \Big \{ -{1/4}, -{1/21}, -{23/1701}, \cdots \Big \} \\
&& f_i = \Big \{ 1, 17/21, 341/567, \cdots, \Big \}
\een
\ees
The power spectrum in modified cosmology will be denoted as ${\tilde P}_{\delta}$ and in the fiducial cosmology by
$P_{\delta}$ are related by the following expression:
\ben
&& P_{\delta}(k|\delta_L) = [1+\delta_{\rho}]{\tilde P}_{\delta}([1+\delta_a]k).
\label{eq:afund0}
\een
The growth-only response function is denoted by $G_n$ and the
total response function by $R_n$.
\bes\ben
&&{\tilde P}_{\delta}(k,t) = \sum_{n =0}^{\infty}
    {\delta_L^n \over n!} G_n(k,t)\, P_{\delta}(k,t); \label{eq:afunda1} \\
&&    {P}_{\delta}(k,t|\delta_L)= \sum_{\alan{n=0}}^{\infty}
    {\delta_L^n \over n!} R_n(k,t)\, P_{\delta}(k,t). \label{eq:afunda2}
\een\ees

The nth-order response function for the density contrast is
given by the nth-order derivative of the power spectrum
with the linearly extrapolated overdensity $\delta_{L0}$.
A normalisation is also introduced by the power spectrum
which renders the response function dimensionless.

\bes\ben
&& R_n(k,t) = {1\over P_{\delta}(k,t)}{d^n P_{\delta}(k,t |\delta_L) \over d\delta^n_L(t)} \Big |_{\delta_L=0}; \\
&& G_n(k,t) = {1\over P_{\delta}(k,t)}{d^n {\tilde P}_{\delta}(k,t) \over d\delta^\rtrv{n}_L(t)} \Big |_{\delta_L=0}. 
\label{eq:funda}
\een\ees
The response functions can be recovered by Taylor expanding
Eq(\ref{eq:afund0}). Next, we will focus on response function
for the power spectrum of the weak lensing convergence.
We will express these response functions in terms of the
3D response function for matter power spectrum.

The 2D power spectrum is given by:
\ben
{\cal{C}}_{\ell}(\kappa_L) =
\int_0^{r_s} dr {\rtrv{W}^2(r) \over  {d^2_A}(r)}
P_{\delta}\left ({\ell \over r}; r \Big |\delta_L \right ).
\label{eq:localps}
\een
We will refer to this as the local angular power spectrum. The
global angular power spectrum is recovered by taking
$\delta_L = 0$ which leads to $\kappa_L=0$ and we recover
Eq.(\ref{eq:global_ps}).
%
%
%
%
%%%%%%%%%%%%%%%%%%%%%%%%%%%%%%%%%%%%%%%%%%%%%%%
\begin{figure}
  \begin{center}
  \begin{minipage}[b]{0.3\textwidth}
    \includegraphics[width=\textwidth]{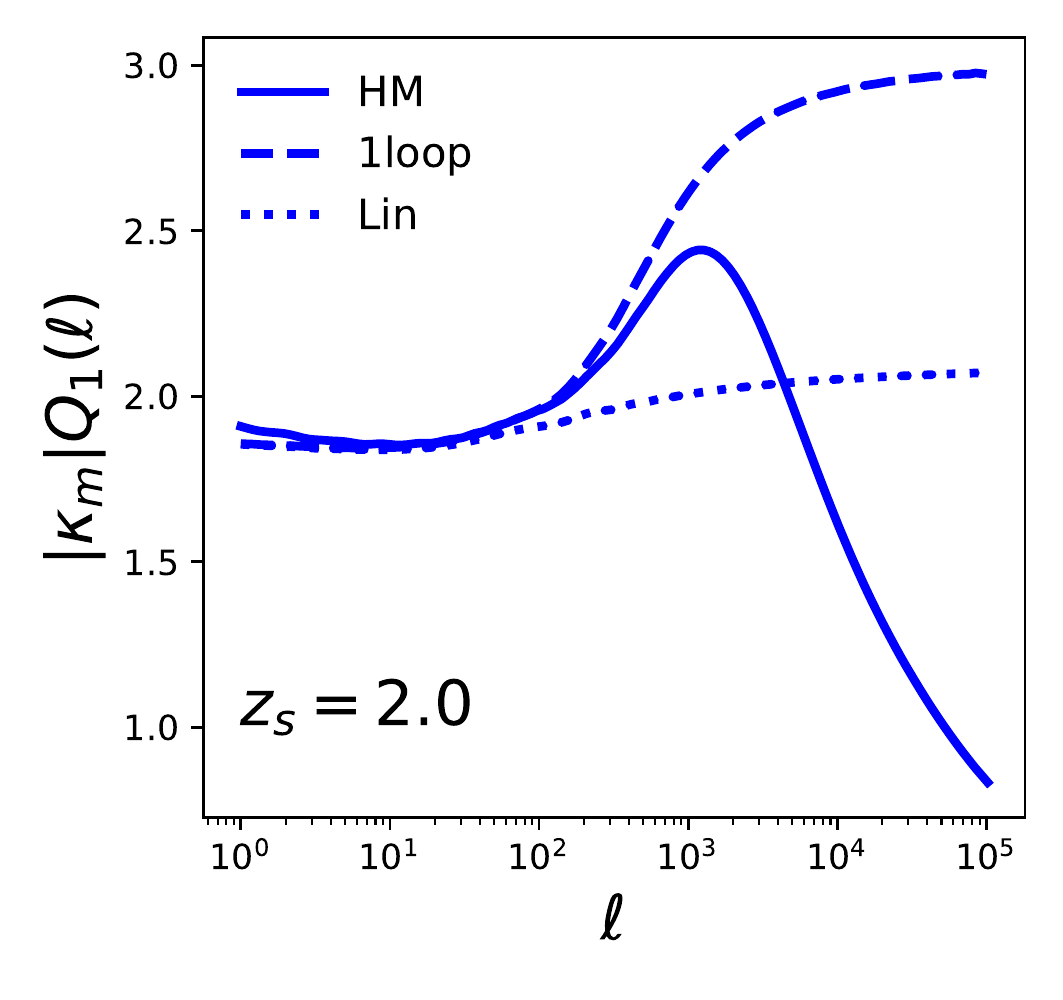}
    %\label{fig:1}
  \end{minipage}
  \begin{minipage}[b]{0.3\textwidth}
    \includegraphics[width=\textwidth]{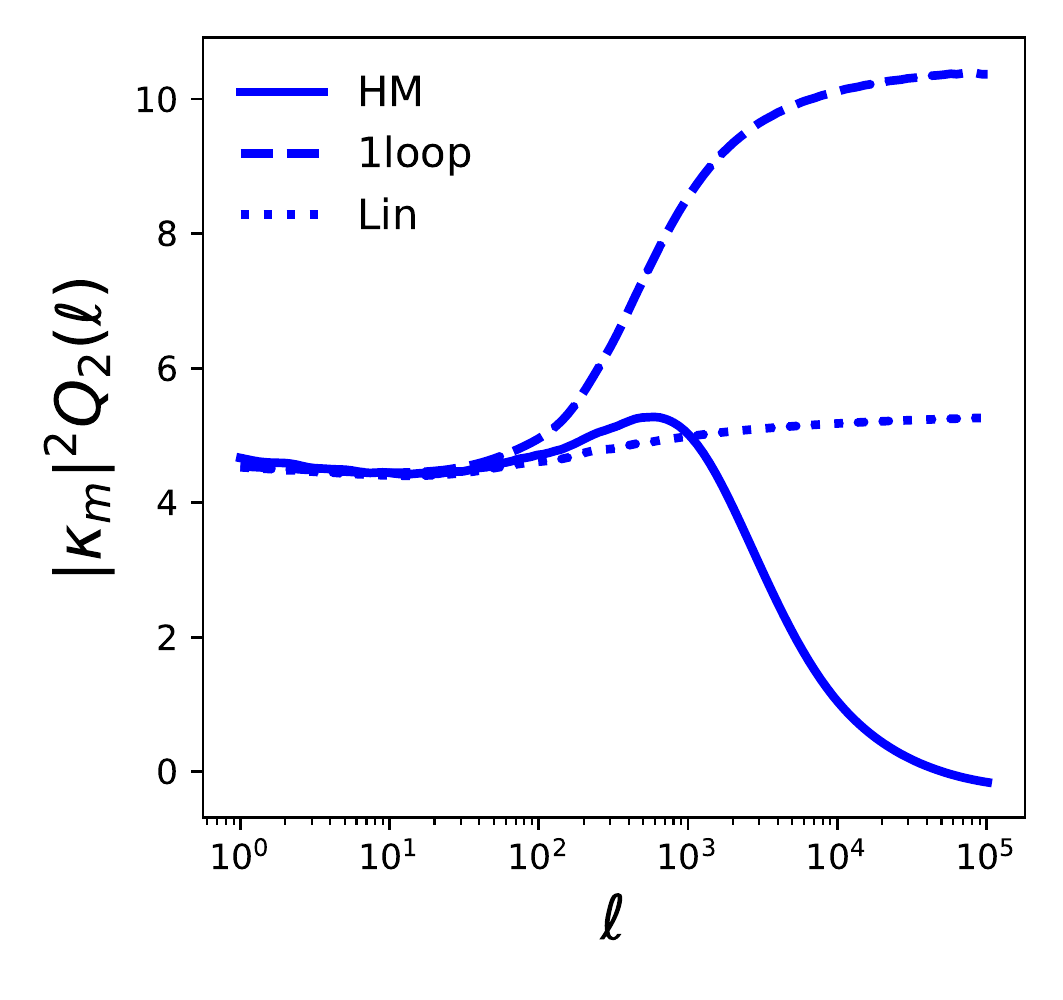}
    %\label{fig:2}
  \end{minipage}
  \begin{minipage}[b]{0.3\textwidth}
    \includegraphics[width=\textwidth]{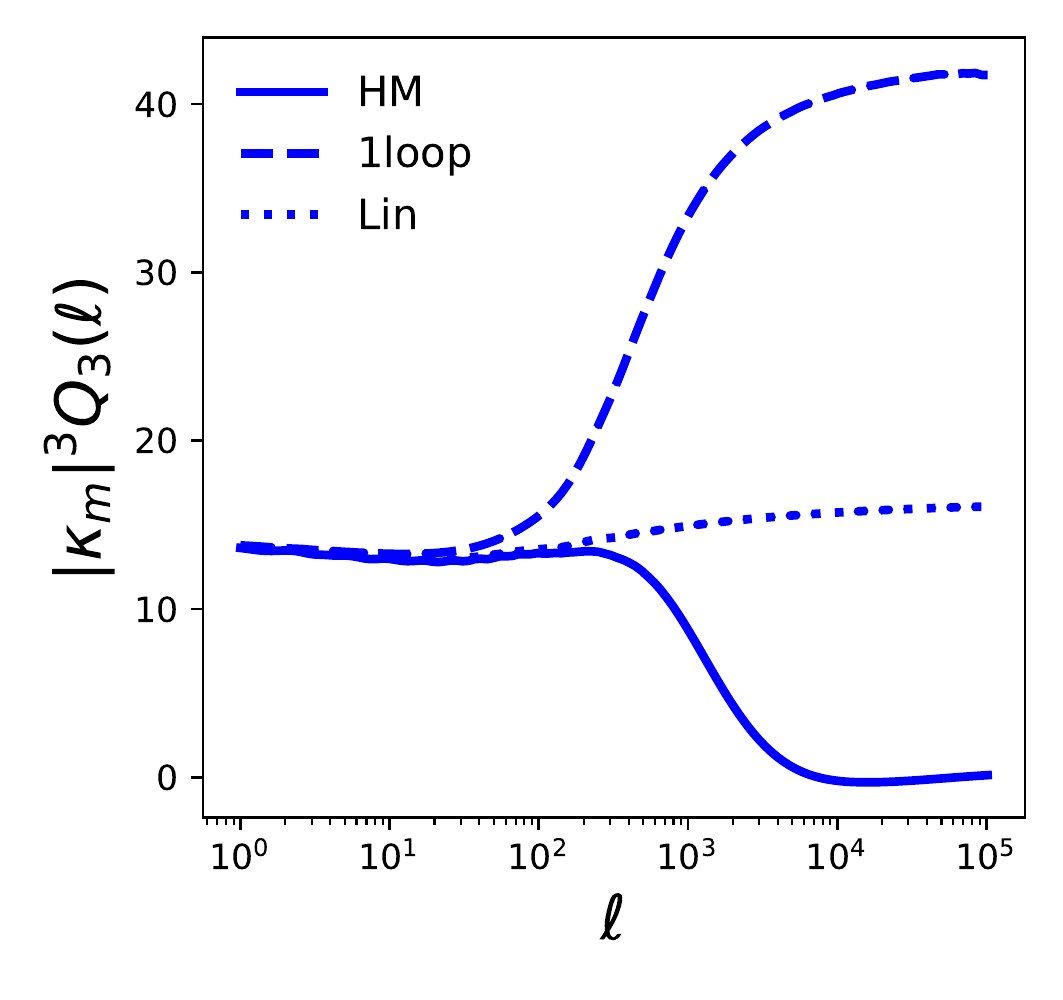}
    %\label{fig:2}
    %\label{fig:QN_z2}
  \end{minipage}
  \caption{Same as Fig.\ref{fig:QN_z1} but for source redshift $z_s=2$.
  }
  \label{fig:QN_z2}
  \end{center}
\end{figure}
%%%%%%%%%%%%%%%%%%%%%%%%%%%%%%%%%%%%%%%%%%%%%%%%%%
%
%
%
Using Eq.(\ref{eq:afunda1}) we can write:
\ben
{\cal C}_{\ell}(\kappa_L)
= \sum_{n=0}^{\infty}{1\over n!} [\delta_{L0}]^n
\int_0^{r_s} dr  {\rtrv{W}^2(r) \over {d^2_A}(r)}
R_n\left ({\ell \over d_A(r)},r \right )[D_+(r)]^n
P_{\delta}\left ({\ell\over d_A(r)},r \right ).
\een
%If we replace $\kappa_L =  \delta_L |\kappa_m|$.
%\ben
Using the following notation:
\ben
&& \kappa_L =  \delta_L |\kappa_m|; \quad \kappa_m = -\int_0^{r_s} dr D_+(r)\, \rtrv{W}(r); 
\label{kappa_min}
\een
where $\delta_L$ is the local over(under)-density and corresponding
projected convergence is $\kappa_L$
\ben
&& {\cal C}_{\ell}(\kappa_L)
= \sum_{n=0}^{\infty}{1\over n!}
\left [{\kappa_L \over |\kappa_m|} \right ]^n
\int_0^{r_s} dr {{W}^2(r) \over {d^2_A}(r)}
R^{}_n\left [{\ell \over d_A(r)},r\right ][D_+(r)]^n
P_{\delta}\left ({\ell\over d_A(r)},r \right ).
\label{eq:response1}
\een
If we define the response functions for the 2D as ${\cal Q}_n(\ell)$ we can write:
\ben
&& {\cal C}_{\ell}(\kappa_L ) =  \sum_{n=0}^{\infty}{1\over n!}\;
{\cal Q}_n(\ell) \; {\kappa^n_L}\; {\cal C}_{\cal \ell}
\label{eq:cls_response}
\een
Comparing Eq.(\ref{eq:cls_response}) with Eq.(\ref{eq:response1}) we
deduce that:
\ben
Q_n(\ell) = {1\over {\cal C}_{\ell}} {1  \over |\kappa_m|^n}
\int_0^{r_{s}} dr {{W}^2(r) \over  {d^2_A}(r)}
R_n\left [{\ell \over d_A(r)},r\right ][D_+(r)]^n
P_{\delta}\left ({\ell\over d_A(r)},r \right ).
\label{eq:def_QN}
\een
The respone function for cross-correlation of two different tomographic bins
denoted by indices $i$ and $j$ can be derived similarly. We start from the definition
of local cross-spectra:
\ben
{\cal{C}}^{ij}_{\ell}(\kappa_{LX}) =
\int_0^{r_s} dr {{W}^i(r){W}^j(r) \over  {d^2_A}(r)}
P_{\delta}\left ({\ell \over r}; r \Big |\delta_L \right )
\quad \; {\kappa_{LX}} \equiv [{\kappa_{im}}{\kappa_{jm}}]^{n/2}
\label{eq:local_cross_Cls}
\een
We will later define the local cross-spectra in terms of
$\kappa_{LX}$ which is a geometric
mean of $\kappa_{mi}$ and $\kappa_{mj}$
i.e. $\kappa_{m}$ of the tomographic bins $i$ and $j$ respectively.
Mathematically,
\ben
&& \kappa_{iL} =  \delta_L |\kappa_{im}|; \quad
\kappa_{im} = -\int_0^{r_s} dr D_+(r)\, W_i(r); 
\label{eqlkappam}
\een
The response functions ${\cal Q}^{ij}_n(\ell)$
for cross-spectra involving
two tomographic is expressed as:
\ben
&& {\cal C}^{ij}_{\ell}(\kappa_{LX} ) =  \sum_{n=0}^{\infty}{1\over n!}\;
    {\cal Q}^{ij}_n(\ell) \; {\kappa^n_{LX}} {\cal C}^{ij}_{\cal \ell};
\label{eq:cls_cross_response}
\een
Combining Eq.(\ref{eq:cls_cross_response}) and Eq.(\ref{eq:local_cross_Cls})
\ben
Q^{ij}_n(\ell) = {1\over {\cal C}^{ij}_{\ell}} {1  \over |\kappa_{LX}|^n}
\int_0^{r_{s}} d\tilde r {{W}^i(r){W}^j(r) \over  {d^2_A}(r)}
R_n\left [{\ell \over d_A(r)},r\right ][D_+(r)]^n
P_{\delta}\left ({\ell\over d_A(r)},r \right ).
\label{eq:def_ijQN}
\een
Throughout, we have used Limber approximation.
The FFTlog based approach is often used to go beyond the Limber approximation
\citep{xFFT} in the modelling of projected power spectrum.
It is possible to incorporate a similar method to model the low-$\ell$
behaviour of the response functions.
The expressions above are derived for single
source plane. For generalisation to a source
distribution specified by $n(z)$ we need to
replace $W(z)$ in Eq.(\ref{eq:localps}), Eq.(\ref{eq:def_QN}) and
other equations by $\hat W(z)$
as defined in Eq.(\ref{eq:define_hat}). The definition
of $\kappa_m$ in Eq.(\ref{kappa_min}) similarly will have to be modified.

Next, from \citep{Thesis} we have  used the following expressions:
\bes                                   
\ben
&& R_0(k) =1; \label{eq:R0} \\
&& R_1(k) = f_1 + e_1 {k P^{\prime}(k)\over P(k)} + G_1(k); \label{eq:R1} \\
&& {1\over 2}R_2(k) = f_2 + e_2 {k P^{\prime}(k) \over P(k)} 
+ e_1^2  {k^2 P^{\rtrv{\prime \prime}}(k) \over 2 P(k)}   
+ {1\over 2} G_2{(k)} + f_1 G_1(k) \nonumber \label{eq:R2} \\
&& \quad  + \rtrv{f_1} e_1 {k P^{\prime}(k) \over P(k)} +
e_1 {k P^{\prime}(k) \over P(k)} G_1(k) + e_1 k G^{\prime}_1(k);
\label{eq:ra}\\
&& {1\over 6}{R_3(k)} = f_1 \rtrv{G_1}(k)e_1 {k P^{\prime}(k) \over P(k)} + f_3 + {\rtrv{G_3}(k)\over 6} + e_3 {k P^{\prime}(k) \over P(k)} +
f_1 {G_2(k) \over 2}
+ f_1e_2{k P^{\prime}(k) \over P(k)} \nonumber \label{eq:R3} \\
&& + f_1e_1^2{k^2 P^{\prime\prime}(k) \over 2 P(k)} + f_2 G_1(k)
+ f_2 e_1 {k P^{\prime}(k) \over P(k)} + 
 (f_1e_1 + e_2) k G_1^\prime(k) + e_1^2 {k^2 G_1^\rtrv{\prime\prime}(k) \over 2}
\nonumber \\
&& + e_1 k {G^{\prime}_2(k)\over 2}
+ e_1^2 {k P^{\prime}(k) \over P(k)} {k G^{\prime}_1(k)}
+ e_1^3{k^3 P^{\prime\prime\prime}(k) \over 6 P(k)}
+ 2 e_1e_2 {k^2 P^{\prime\prime}(k) \over 2 P(k)}
\nonumber \\
&& + e_1 {k P^{\prime}(k) \over P(k) }{G_2(k) \over 2 } 
+  G_1(k) \left ( e_2{k P^{\prime}(k) \over P(k) }
+ e_1^2 {k^2 P^{\prime\prime}(k) \over 2 P(k) } \right ). \label{eq:R4}
%\label{eq:rc}
\een
\ees
In our notation 
$\hat D_+(t)$ represents the linear growth rate.
The growth only response functions, denoted as $G_n$, takes a particularly simpler form when
computed using linear theory.

Notice that the response functions ${R_n(k)}$ take a rather simpler
form when we approximate the power spectrum locally as a power law.
In this case, we can write $P(k)\propto k^n$ which leads us to
$kP^{\prime}(k)/P(k) = n$ and $k^2\,P^{\prime\prime}(k)/P(k) = n(n-1)$,
and to a good approximation the growth rate $G(k)$ is scale-independent.
%%%%%%%%%%%%%%%%%%%%%%%%%%%%%%%%%%%%%%%
\begin{figure}
  \begin{center}
  \begin{minipage}[b]{0.3\textwidth}
    \includegraphics[width=\textwidth]{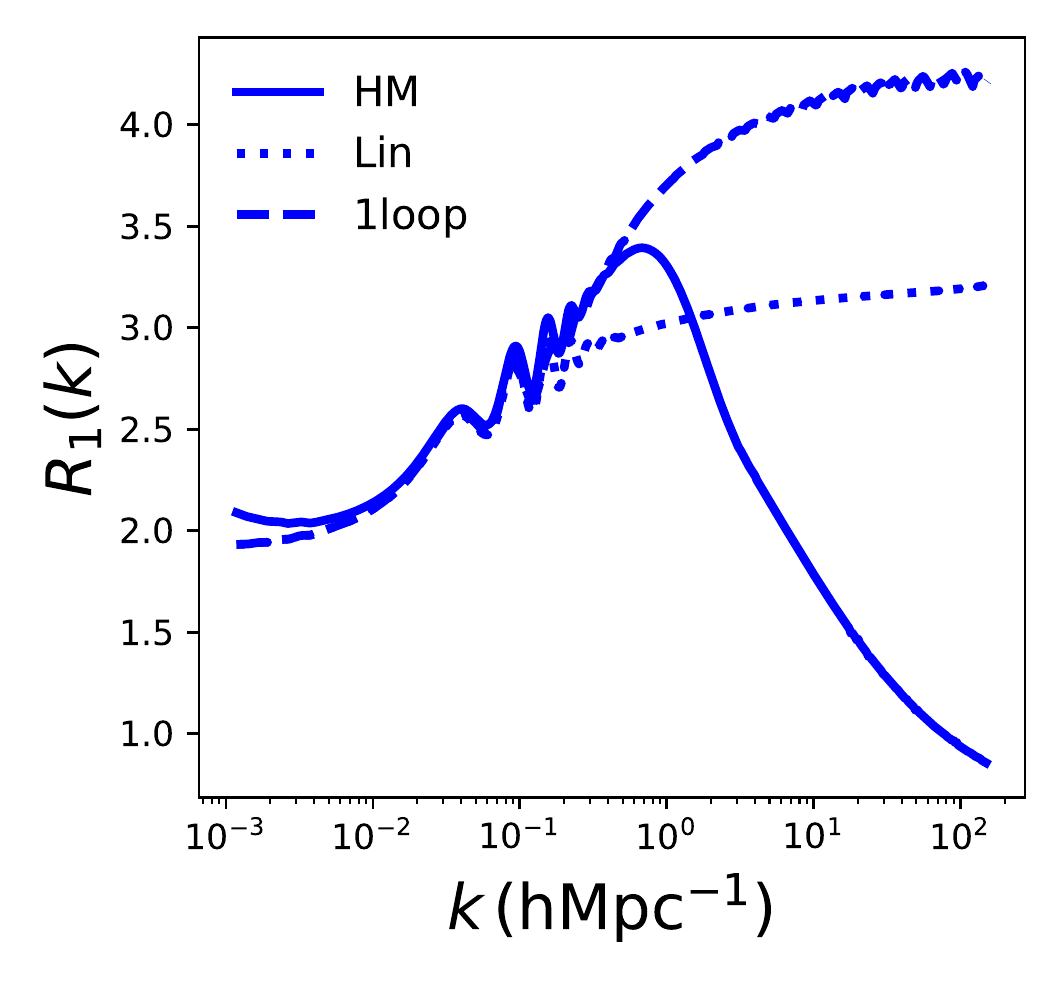}
  \end{minipage}
  \begin{minipage}[b]{0.3\textwidth}
    \includegraphics[width=\textwidth]{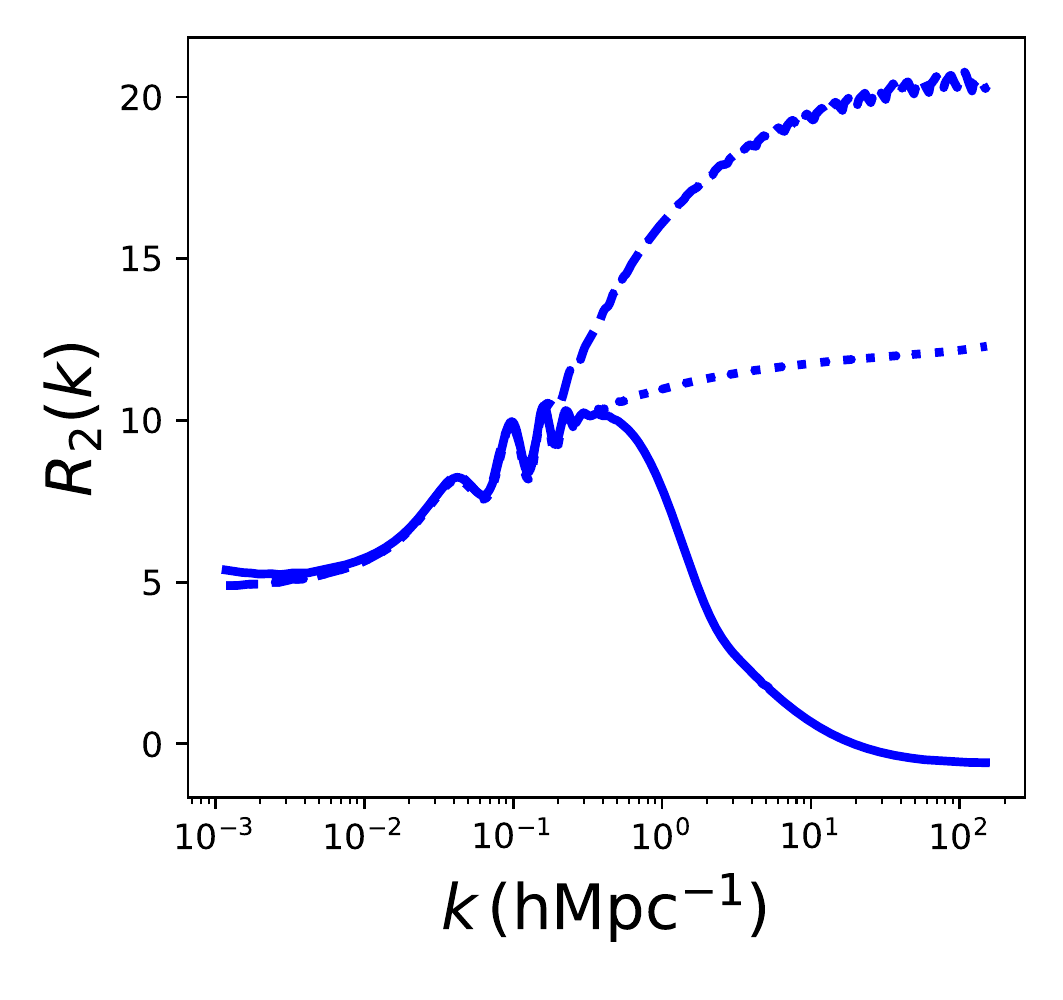}
  \end{minipage}
  \begin{minipage}[b]{0.3\textwidth}
    \includegraphics[width=\textwidth]{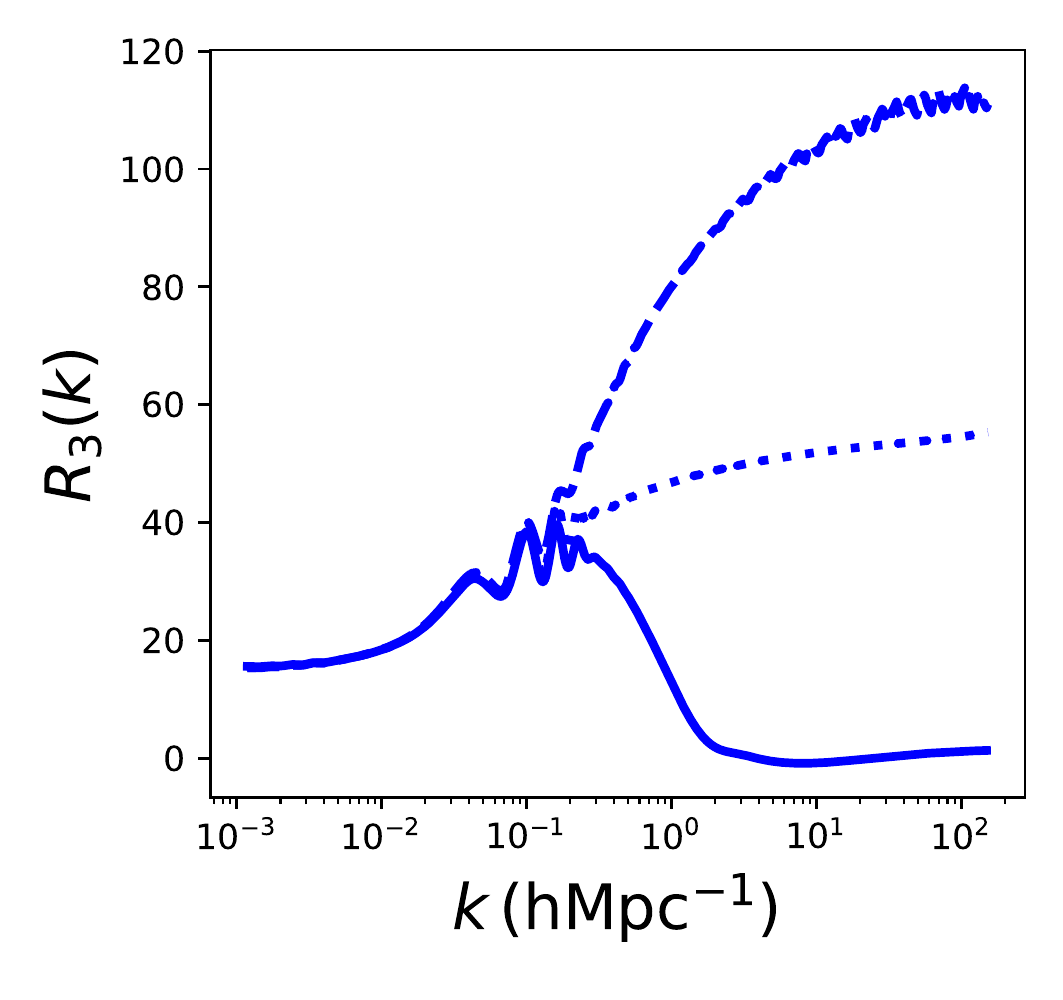}
  \end{minipage}
  \caption{The response functions $R_N$ defined in
    Eq.(\ref{eq:R1})-Eq.(\ref{eq:R4}) are shown. From
    left to right panels depict $N=1,2$ and $N=3$. The simulation redshift is at $z_s=1$.
    Various line styles correspond to different analytical
    models, linear, one-loop and halo model as indicated (see text for details).}
  \label{fig:R1}
  \end{center}
\end{figure}
%%%%%%%%%%%%%%%%%%%%%%%%%%%%%%%%%%%%%%%%
%
\bes\ben
&& G_n = {1 \over D_+^2} {d^n \rtrv{\tilde D}_+^2 \over d \delta^n_L} \Big |_{\delta_L=0}; \\
&& \tilde D_+(t) = D_+(t) \rtrv{\sum_{n=0}^{\infty}} \rtrv{g}_n [\delta_L]^n;  \\
&& \rtrv{g_{n=0, 1,2,3,4} = \left \{1, {13/21}, {71/189}, \cdots \right \};}  \\
&& G_{n=0, 1,2,3,4} = \left \{1, {26/21}, {3002/1323}, \cdots \right \}.
\een\ees
In the linear theory Eq.(\ref{eq:afund0}) takes the following form for
the linear power spectrum local $P_{L}$ and the fiducial power spectrum $P_{lin,fid}$:
\ben
P_{L}(k,t |\delta_L ) = (1 + \delta_\rho(t))
\left ( {\tilde D_+(t) \over D_+(t)} \right )^\rtrv{2}
{\tilde P}_{L,fid}(k,t ).
\een
Next, we turn our attention to a perturbative quasilinear 
calculation of response functions.
The response functions for the density contrast $\delta$ are related to the angle averaged squeezed limits of correlation functions $S^{\delta}_{N-2}$ defined as follows:
\bes
\ben
&& S^{\delta}_{N-2}(k,k^{\prime}, k_1, \cdots k_{N-2})
=  \int {d\oh_1 \over 4\pi}\cdots\int {d\oh_{N-2} \over 4\pi} 
\langle \delta(\bk)\delta(\bk^{\prime})\delta(\bk_1)\cdots\delta(\bk_{N-2})\rangle^{\prime}. \\
&& R_{N-2}(k) = \lim_{k_i \rightarrow 0} {S_{N-2}(k,k^{\prime}, k_1, \cdots k_{N-2})
  \over P^{}_L(k_1) \cdots P^{}_L(k_{N-2})}
\een
\ees
Here, $d\hat\Omega_i=\sin\theta_id\phi_i$. The angles
$\hat\Omega_i= (\theta_i,\phi_i)$ are the angles associated 
with wave vector${\bf k}$ and $P_L$ is the linear power spectrum.
We have used following shorthand notation above:
\ben
\langle \delta(\bk)\delta(\bk^{\prime})\delta(\bk_1)\cdots\delta(\bk_{N-2})\rangle^{\prime}
= (2\pi)^3 \delta_{\rm 2D}(\bk_1)+\cdots+\bk_{N-2}) \nn 
\,\times\,\langle \delta(\bk)\delta(\bk^{\prime})\delta(\bk_1)\cdots\delta(\bk_{N-2})\rangle
\een
Here, $\delta_{\rm nD}$ represents the $n$-dimensional Dirac delta function.
In case of 2D convergence or $\kappa$similarly we have:
\bes\ben
&& S^{\kappa}_{N-2}(\ell,\ell^{\prime}, \ell_1, \cdots \ell_{N-2}) = \int {d\theta_1 \over 2\pi}\cdots\int {d\theta_1 \over 2\pi} 
\langle \kappa(\bl)\kappa(\bl^{\prime})\kappa(\bl_1)\cdots\kappa(\bl_{N-2})\rangle^{\prime}. \\
&&  Q^{}_{N-2}(\ell) =
     \lim_{\;\ell_i \rightarrow 0} {S^{\kappa}_{N-2}(\ell,\ell^{\prime}, \ell_1, \cdots \ell_{N-2})
          \over {\cal C}^{\kappa}_{\ell_1} \cdots {\cal C}^{\kappa}_{\ell_{N-2}}}.
     \een
In the above expression $\theta_i$ is the polar angle of the vector $\bl_i$ and $\ell_i=|\bl_i|$. The power spectrum 
${\cal C}^{\kappa}_{\ell_1}$ is the linear convergence power spectrum for the
same source distribution.
The following notation was used:
     \ben
&& \langle \kappa(\bl)\kappa(\bl^{\prime})\kappa(\bl_1)\cdots\kappa(\bl_{N-2})\rangle^{\prime}=  
(2\pi)^2 \delta_{\rm 2D}(\bl_1+\cdots+\bl_{N-2}) \nn && \quad\quad\quad\quad
\quad\quad\quad\quad\quad\quad\,\times\,\langle\kappa(\bl)\kappa(\bl^{\prime})\kappa(\bl_1)\cdots\kappa(\bl_{N-2})\rangle^{}
\een
\ees
     The projected higher-order squeezed spectra are not identical to
     the projected response functions of same order. See for exact perturbative
     results in Appendix-\textsection\ref{sec:perturbative}
\begin{figure}
  \begin{center}
  \begin{minipage}[b]{0.3\textwidth}
    \includegraphics[width=\textwidth]{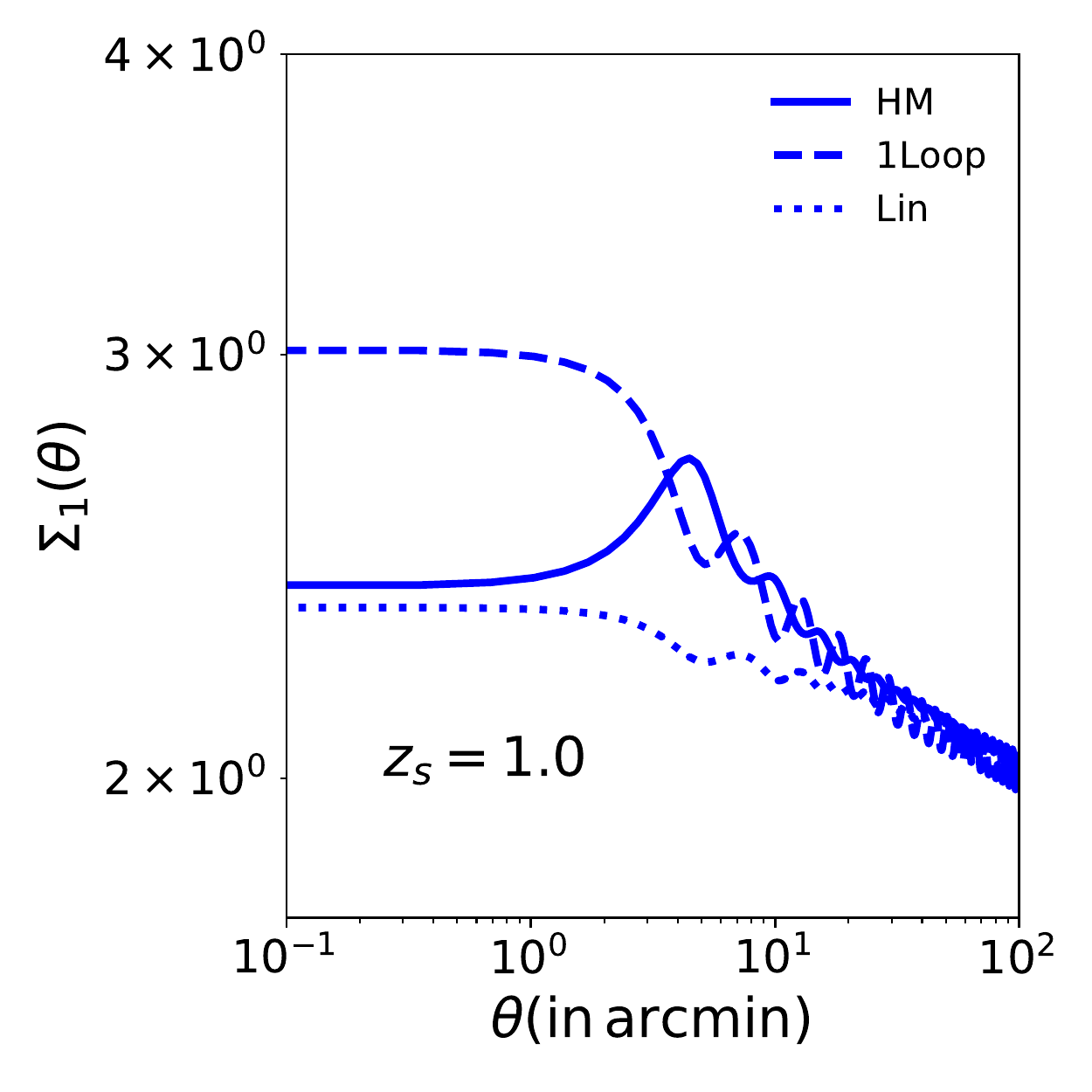}
    %\label{fig:1}
  \end{minipage}
  \begin{minipage}[b]{0.3\textwidth}
    \includegraphics[width=\textwidth]{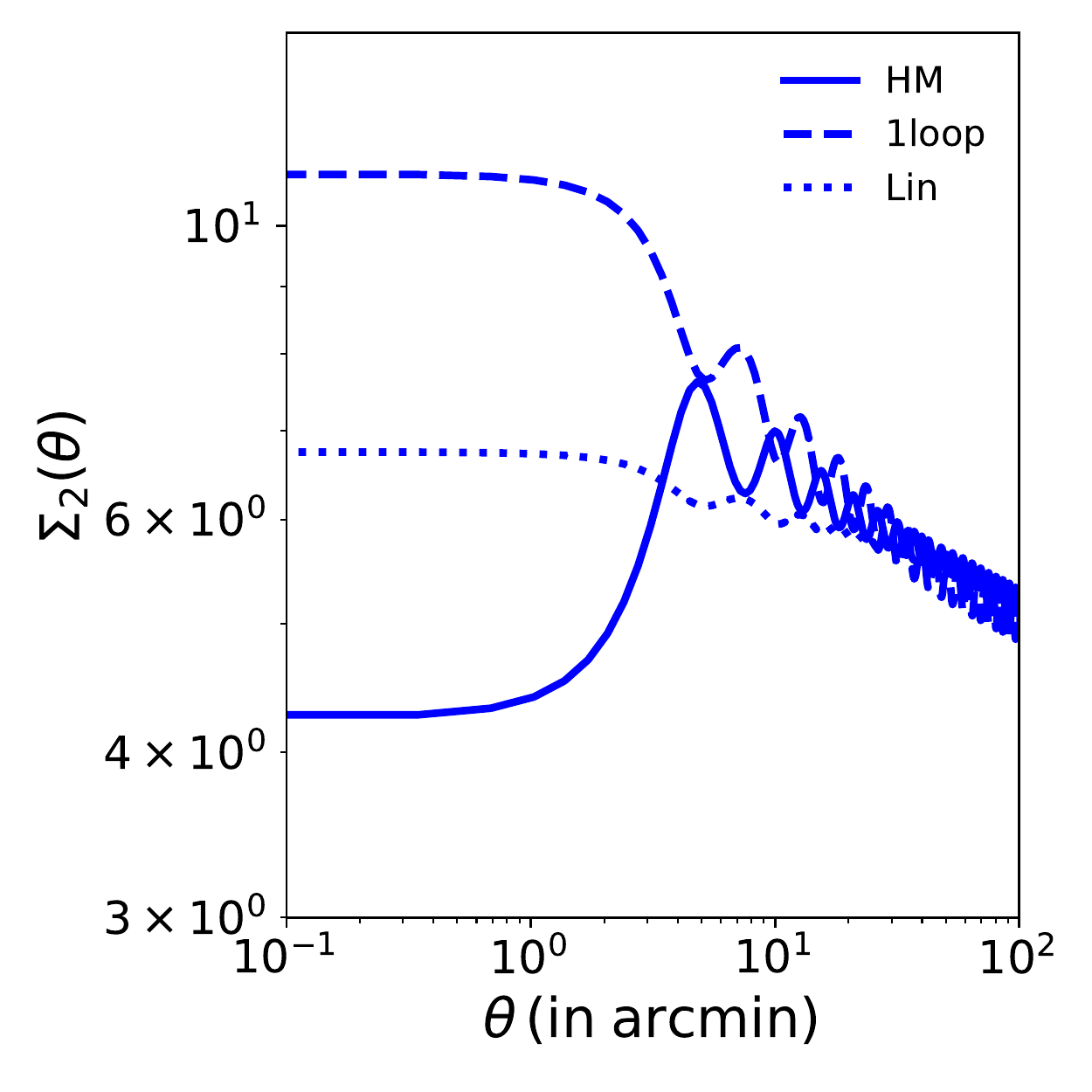}
    %\label{fig:2}
  \end{minipage}
  \begin{minipage}[b]{0.3\textwidth}
    \includegraphics[width=\textwidth]{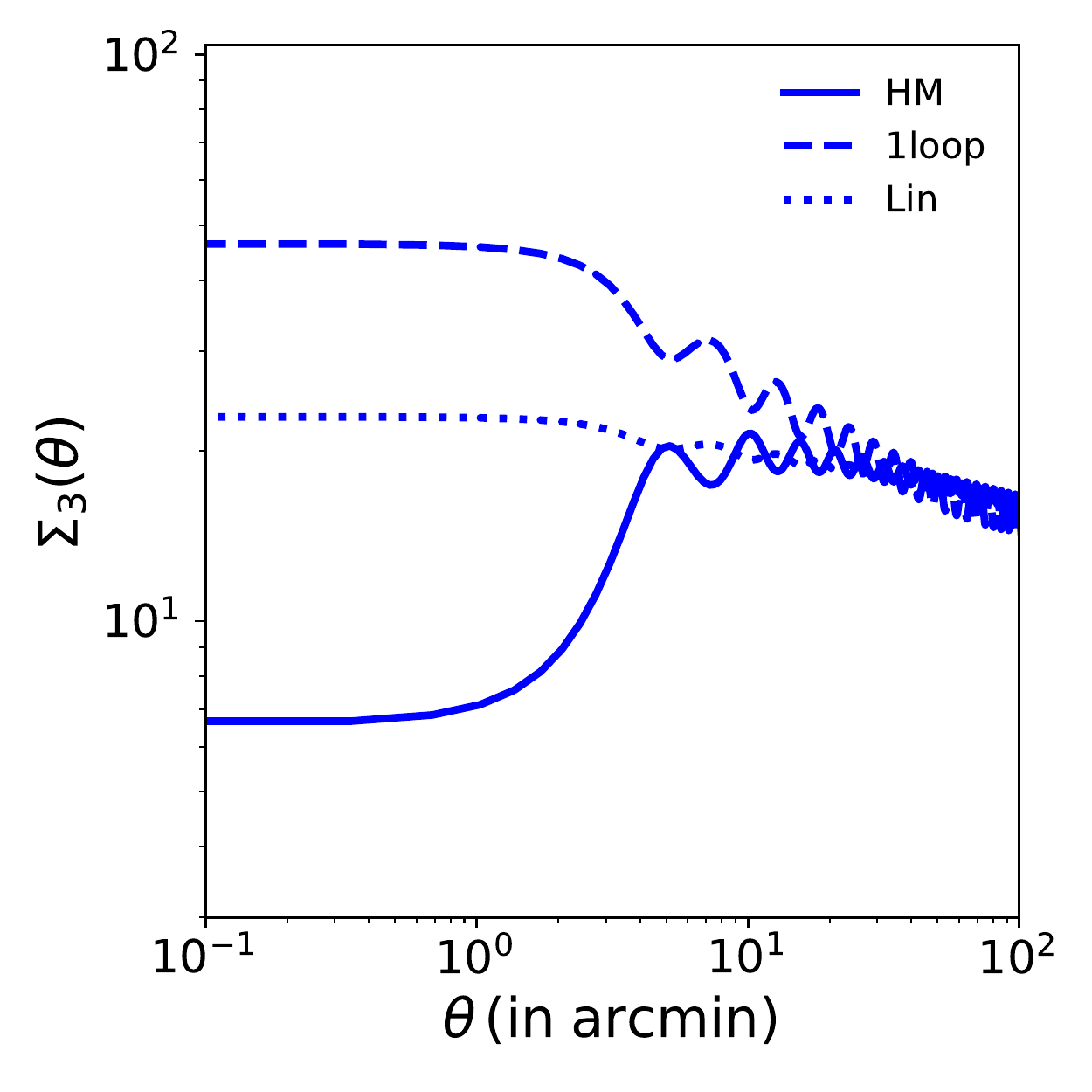}
    %\label{fig:2}
  \end{minipage}
  \caption{The response functions $\Sigma_N$ defined in  Eq.(\ref{eq:response})
  are shown for the source redshift $z_s=1.0$. From
    left to right panels depict $N=1,2$ and $N=3$. 
    Various line styles correspond to different analytical
    models, linear, one-loop and halo model as indicated.}
  \label{fig:QN_corr}
  \end{center}
\end{figure}
%
     
%
%%%%%%%%%%%%%%%%%%%%%%%%%%%%%%%%%%%%%%%%%%%%%%%%%
\subsection{Beyond Linear Theory - Loop Corrections}
\label{subsec:pt}
%%%%%%%%%%%%%%%%%%%%%%%%%%%%%%%%%%%%%%%%%%%%%%%%%

In the standard perturbation theory (SPT) the power spectrum
at a redshift $z$ has the following expression \rtrv{(e.g., \cite{bernardeaureview})}:
\ben
P^{\rm SPT}(k) = D_+^2 P_{L}(k) +  D_+^4 P^{\rm{1-loop}}_{}(k)
+ D_+^6 P^{\rm{2-loop}}_{}(k) + \cdots
\label{eq:loop}
\een
Here $P^{\rm{1-loop}}$ $P^{\rm{2-loop}}$ denote loop-level corrections to the linear
power spectrum $P_{L}$. In the previous section we used the
linear power spectrum for computing the response functions. Perturbative corrections
to the linear theory as given in Eq.(\ref{eq:loop}) can improve
these predictions when used in association with Eq.(\ref{eq:ra})-Eq.(\ref{eq:R4}).

%%%%%%%%%%%%%%%%%%%%%%%%%%%%%%%%%%%%%%%%%%
\subsection{Halo Model Response Functions}
\label{sec:hm}
%%%%%%%%%%%%%%%%%%%%%%%%%%%%%%%%%%%%%%%%%%%
We introduce the following notation:
\ben
I_m^n(k_1,\cdots,k_m) \equiv \int d\ln M n(\ln M) \left ( {M\over \bar\rho} \right )^\rtrv{m}
b_n(M) u(M|k_1)\cdots  u(M|k_m).
\een
Here $b_n(M)$ is the $n$-th order bias and $u(M|k_1)$ is the Fourier
transform of the halo profile and $k_i$ are the wave numbers.
The power spectrum in the halo model has two contributions
known as the 1-halo and 2-halo contribution $P_{1h}(k)$ and $P_{1h}(k)$
\citep{halo_review}:
\bes
\ben
&& P_{\rm HM}(k) = P_{2h}(k) + P_{1h}(k); \\
&& P_{2h}(k) = [I^{1}_{1}(k)]^2P_{\rm L}(k); \quad P_{1h}(k) = I_2^0(k,k).
\een
\ees

Following Ref.(\citep{Wagner}) we can write the position-dependent
1-halo contribution as \citep{halo_review}:
\bes
\ben
&& P_{1h}(k,t | \delta_{L0}) =
\sum_{n=0}^{\infty} {1\over n!} I_2^n(k,k,t) \delL.
\een
\ees
The 2-halo contribution is similarly given by \citep{halo_review}:
\ben
&& P^{2h}(k,t,|\delta_{L0}) =\deltaf\delL  \nn
&& \times  \left ( \sum^{\infty}_{n=0}{1\over n! } \rtrv{I_1^{n+1} (k,t)} [\delta_{L}]^n \right ) %\In 
P_{lin}\left [ \left (\sum^{\infty}_{n=0} {1 \over n!} \rtrv{e_n} [\delta_{L}]^n \right )  %\deltaf 
k \right ].
\een
%
%
%\ees
Finally, the response functions in the halo model are given by \citep{Thesis}:
\bes
\ben
&& R_1^{\rm HM}(k) =  \Big [ f_1 + 2g_1 + e_1{d\ln P(k)\over d\ln k} \Big ]
P_{2h}(k,t)+I^1_2(k,k,t); \\
&& R_2^{\rm HM}(k) =  \Big [ 2f_2 + 2f_1g_1 + (f_1+2g_1 \rtrv{) e_1}{d\ln P(k)\over d\ln k} \nonumber \\
&& \quad + 2g_1^2 + 4g_2 + 2e_2 {d\ln P(k)\over d\ln k} +
e_1^\rtrv{2} {1\over P} {d^2\ln P(k)\over d(\ln k)^2 } \Big ] 
P_{2h}(k,t) + I^2_2(k,k,t).
\een
\ees
The $n-$th order response function we have derived correspond to the
Lagrangian density contrast $\delta_a$ and is generally 
denoted as $R^L_n$ to distinguish it from the Eulerian 
response function related to density contrast $\delta_\rho$.
The conversion between the Lagrangian and Eulerian 
response functions is given by:
\bes
\ben
&& R_1^E(k) = R^L_1(k); \\
&& R_2^E(k) = R_2^L(k) − 2f_2R^L_1(k); \\
&& R_3^E(k) = R_3^L(k) − 6f_2R^L_2(k) + 6(2f_2^2 − f_3)R_1(k).
\een
\ees
We have computed the Lagrangian projected 
response functions $Q_N$ by using 
the 3D Lagrangian response function $R^L_1(k)$.
However, the projected Eulerian response functions
can be computed using $R^E_1(k)$ in a straight
forward manner.
%%%%%%%%%%%%%%%%%%%%%%%%%%%%%%%%%%
\begin{figure}
  \begin{center}
  \begin{minipage}[b]{0.3\textwidth}
    \includegraphics[width=\textwidth]{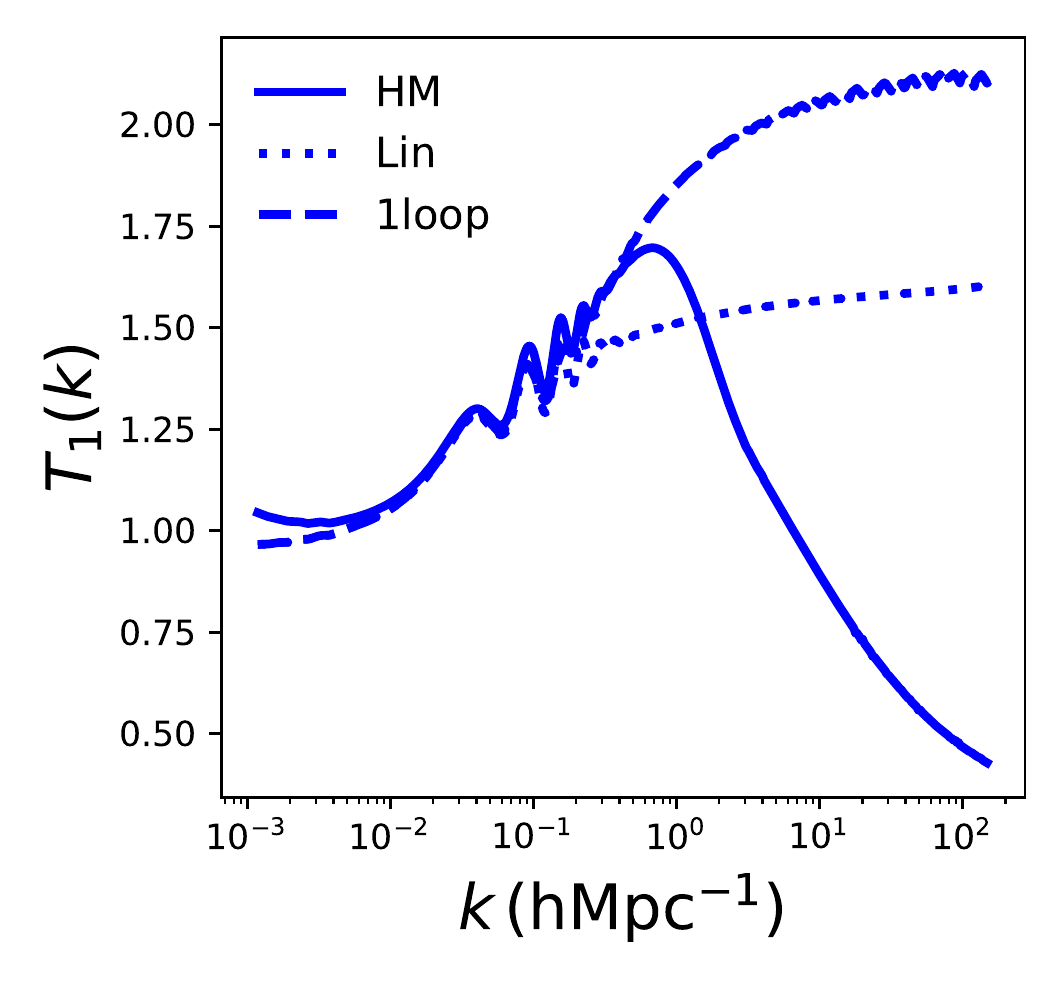}
      %\label{fig:1}
  \end{minipage}
  \begin{minipage}[b]{0.3\textwidth}
    \includegraphics[width=\textwidth]{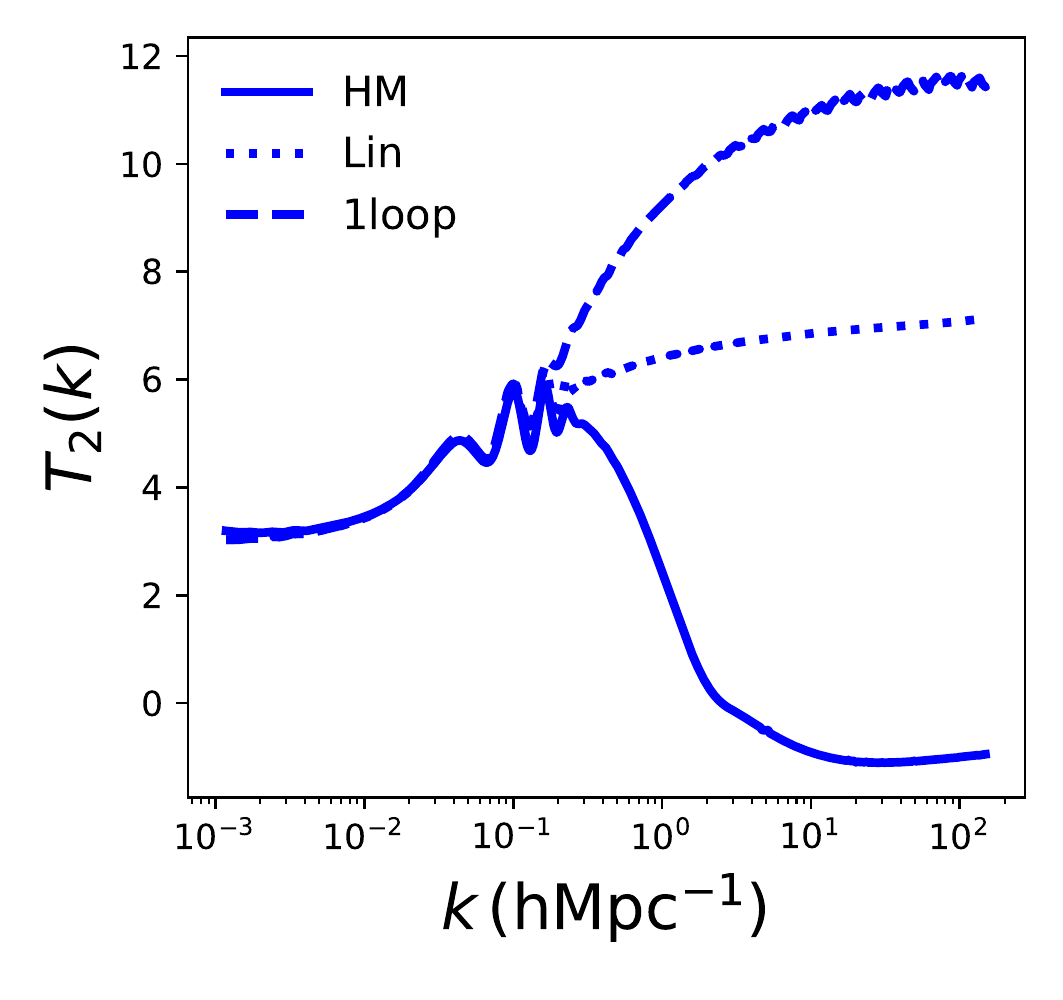}
    %\label{fig:2}
  \end{minipage}
  \begin{minipage}[b]{0.3\textwidth}
    \includegraphics[width=\textwidth]{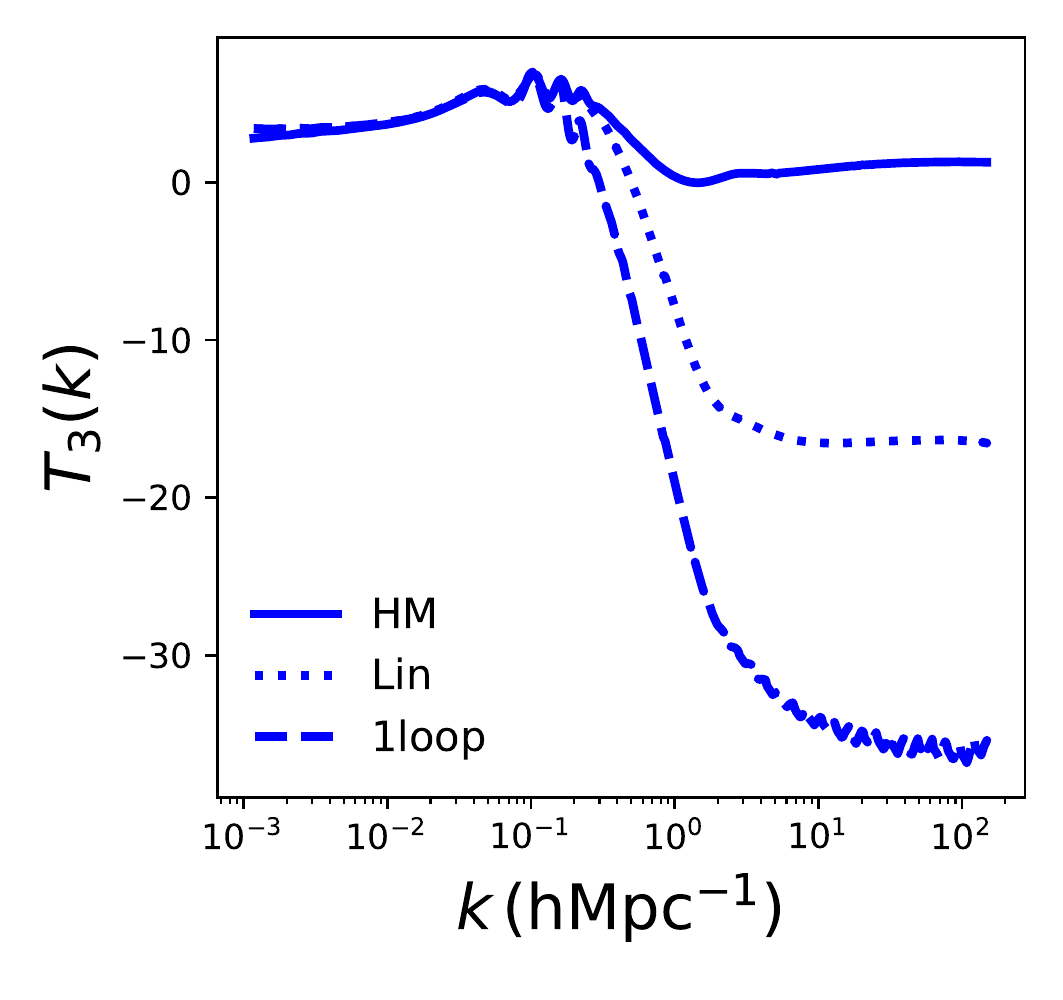}
    %\label{fig:2}
   %\label{fig:QN_z2}
  \end{minipage}
  \caption{The parameters $T_N$ for redshift $z=0$ defined in Eq.(\ref{eq:T1}) -  Eq.(\ref{eq:T3})
    are plotted as a function the wave number $k$. From 
    left to right panels depict $N=1,2,3$. 
    Various line styles correspond to different analytical
    models, linear, one-loop and halo model as indicated (see text for details).}
  \label{fig:TN_z2}
 \end{center}
\end{figure}
%%%%%%%%%%%%%%%%%%%%%%%%%%%%%%%%
%
In addition to the perturbation theory and halo model based results,
the response functions can also be computed using
the Effective Field Theory (EFT) predictions for the
power spectrum \citep{EFT}.

%%%%%%%%%%%%%%%%%%%%%%%%%%%%%%%%%%%%%%%%%%%%%%%%%%%%%%%%%%
\subsection{Response Functions for Two-point Correlation Function}
%%%%%%%%%%%%%%%%%%%%%%%%%%%%%%%%%%%%%%%%%%%%%%%%%%%%%%%%%%
%\cite{Chiang2pt}
%\cite{My2pt}
%\citep{Halder}
In this section we will extend the results derived above to real space
and derive the response functions for the two-point correlation function:
$\xi(\theta_{12})\equiv \langle \kappa({\bm\theta})\kappa(\bm{\theta+\theta_{12}}) \rangle $. Isotropy and homogeneity dictates $\xi(\theta_{12})$ only
depends on the separation $\theta_{12}=|\bm{\theta_{12}}|$.
For surveys with small sky-coverage and
masks with non-trivial topology, response functions defined
for two-point correlation function are easier to implement
\citep{Chiang2pt,My2pt,Halder1}. We begin by defining the local
estimate of the two-point correlation function (2PCF)
$\xi(\theta_{12})$ as
\bes
\ben
&& \xi(\theta_{12}|\kappa_L) = {1\over 4\pi} \sum_{\ell=0}^{\ell_{max}} (2\ell+1)
P_{\ell}(\cos\theta_{12}){\cal C}_{\ell}(\kappa_L).
\label{eq:localcorr}
\een
Here $P_{\ell}$ is the Legendre polynomial of order $\ell$.
The corresponding global two-point correlation function
$\xi(\theta)$ can be recovered
by replacing the local power spectrum ${\cal C}_{\ell}(\kappa_L)$
with its global counterpart ${\cal C}_{\ell}$.
The n-th order response function $\Sigma_n$
as the n- order derivative of the local correlation function
$\xi(\theta |\kappa_L)$ w.r.t the local convergence $\kappa_L$:
\ben
&& \Sigma_n(\theta_{12}) = {1\over \xi(\theta_{12})}{d^n\xi(\theta_{12} |\kappa_L) \over d\kappa^n_L}.
\een
Using Eq.(\ref{eq:cls_response}) in combination
with Eq.(\ref{eq:localcorr}) 
we can express $\Sigma_n(\theta_{12})$ in terms of ${\cal Q}_n(\ell)$
which both carry equivalent information:
\ben
&& \Sigma_n(\theta_{12}) = {1\over \xi(\theta_{12})}
{1\over 4\pi}\sum_{\ell=0}^{\ell_{max}} (2\ell+1) {\cal C}_{\cal \ell}
{\cal Q}_n(\ell)P_{\ell}(\cos\theta_{12}).
\label{eq:response}
\een
\ees
\noindent
The lowest-order response function for two-point correlation functions was studied recently in \citep{Barreira2018,RT2019,My2pt,Halder1,Halder2}.
  
%%%%%%%%%%%%%%%%%%%%%%%%%%%%%%%%%%%%%%%%%%%%%%%%%%%%%%%
\section{Response Functions for 3D Weak Lensing}
\label{sec:3D}
%%%%%%%%%%%%%%%%%%%%%%%%%%%%%%%%%%%%%%%%%%%%%%%%%%%%%%%

A method to use photometric redshifts to study three-dimensional
weak lensing was introduced in \citep{Heavens03}. Subsequently,
this technique was developed by many authors
see, e.g.,\citep{CastroHeavensKitching, photoz}. Here we generalise the
concept of global 3D shear power spectrum to a local one.
%
%
%%%%%%%%%%%%%%%%%%%%%%%%%%%%%%%%%%%%%
\begin{figure}
  \begin{center}
  \begin{minipage}[b]{0.3\textwidth}
    \includegraphics[width=\textwidth]{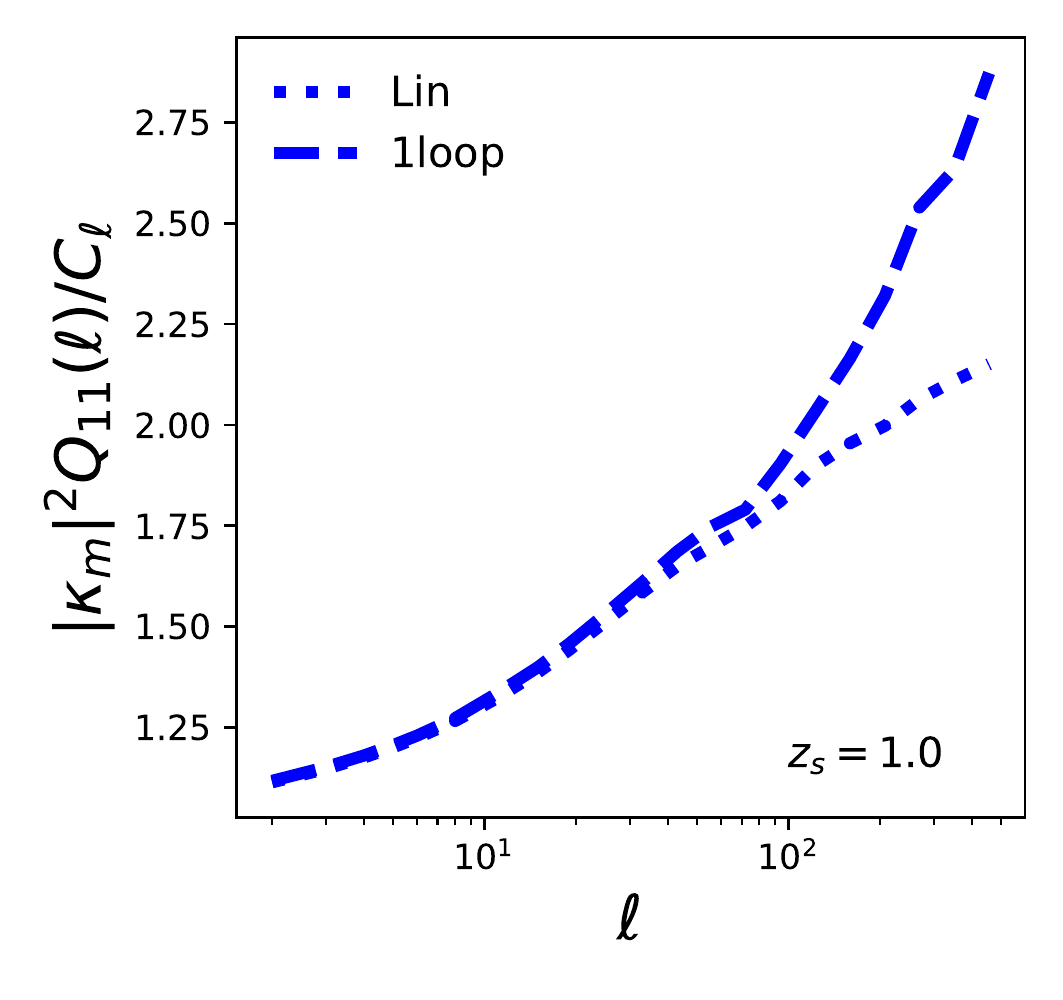}
  \end{minipage}
  \begin{minipage}[b]{0.3\textwidth}
    \includegraphics[width=\textwidth]{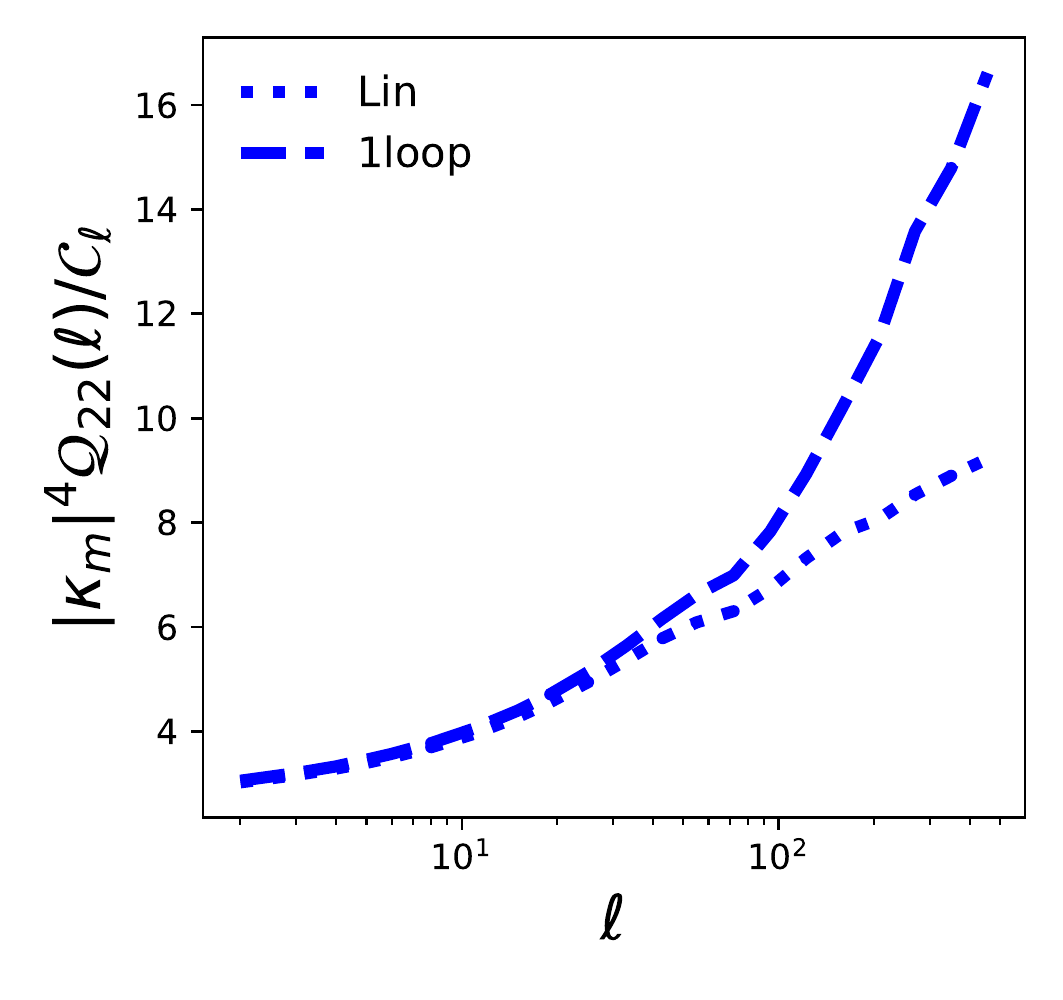}
  \end{minipage}
  \begin{minipage}[b]{0.3\textwidth}
    \includegraphics[width=\textwidth]{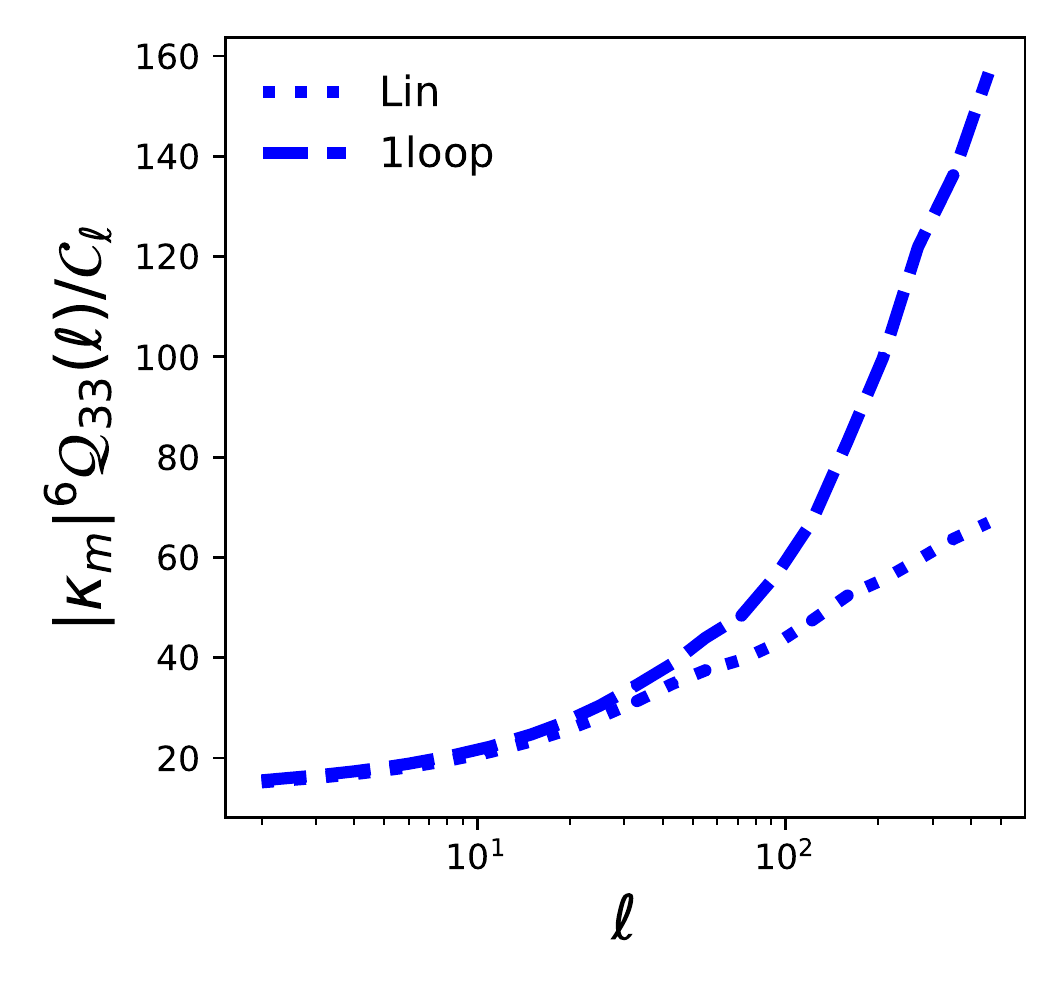}
  \end{minipage}
  \caption{The 3D response functions $|\kappa_m|^{\alpha+\beta}Q_{\alpha\beta}$ defined in  Eq.(\ref{3D_response})
    are shown for the source redshift $z_s=1.0$.
    From left to right panels depict $|\kappa_m|^2Q_{11}$, $|\kappa_m|^4Q_{22}$ and $|\kappa_m|^6Q_{33}$. 
    Various line styles correspond to different analytical
    models, linear (dashed), one-loop (dotted) as indicated.    }
  \label{fig:3D_response}
  \end{center}
\end{figure}
%%%%%%%%%%%%%%%%%%%%%%%%%%%%%%%%%%
%
%%%%%%%%%%%%%%%%%%%%%%%%%%%%%%%%%%%%
\begin{figure}
  \begin{center}
  \begin{minipage}[b]{0.3\textwidth}
    \includegraphics[width=\textwidth]{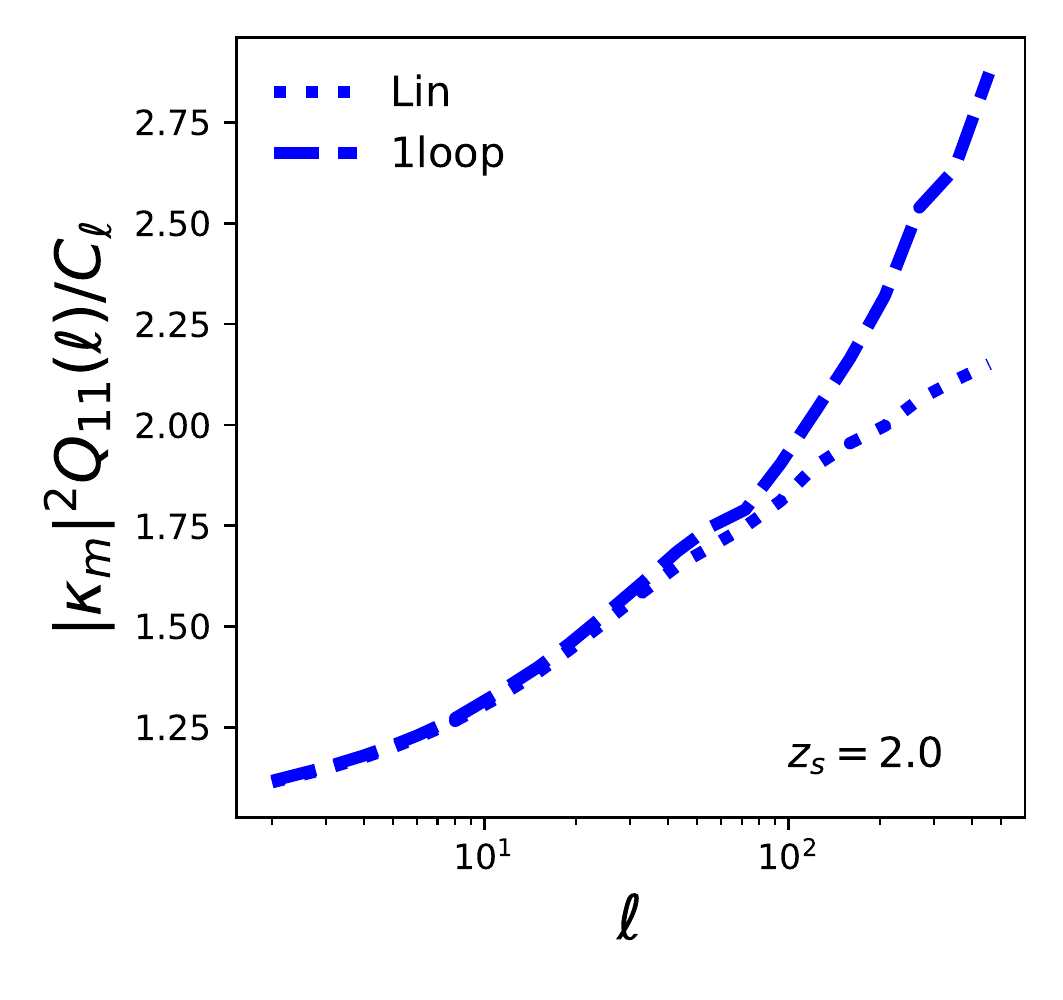}
  \end{minipage}
  \begin{minipage}[b]{0.3\textwidth}
    \includegraphics[width=\textwidth]{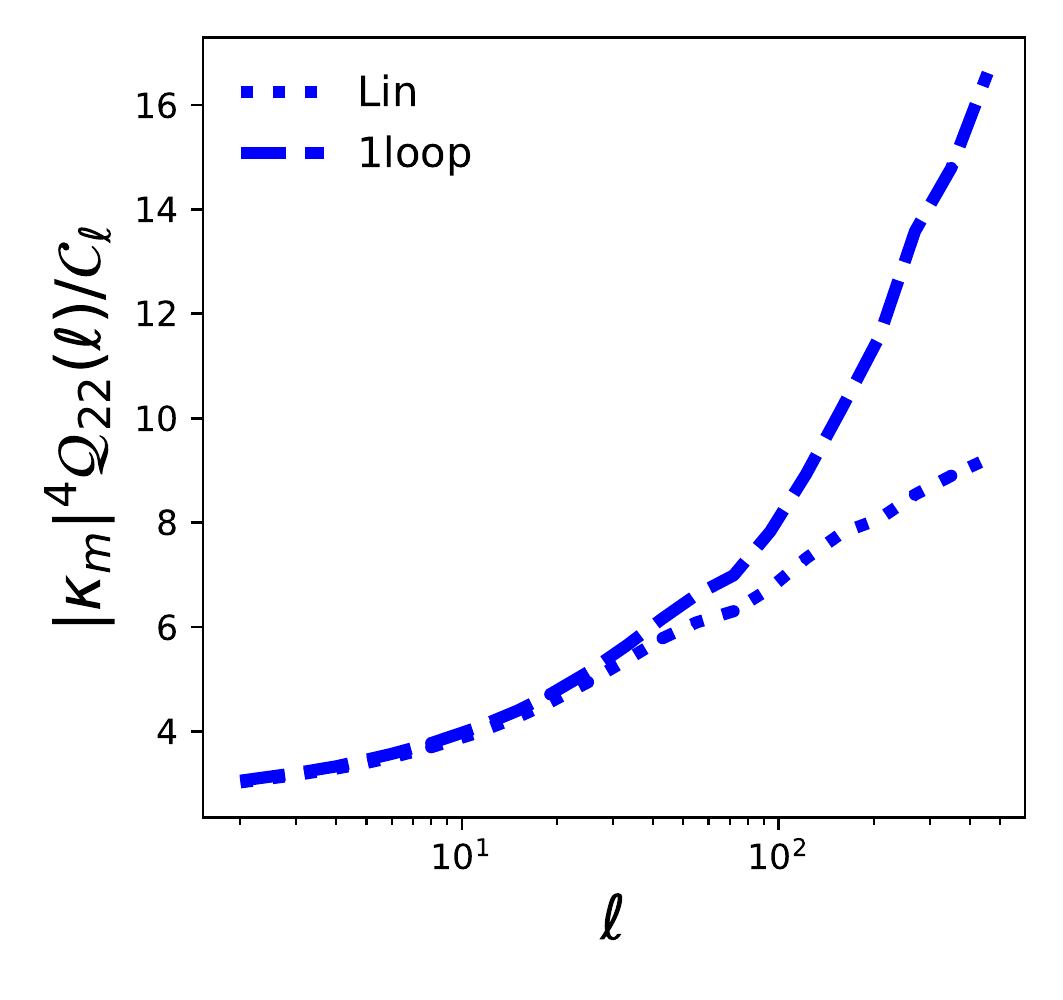}
  \end{minipage}
  \begin{minipage}[b]{0.3\textwidth}
    \includegraphics[width=\textwidth]{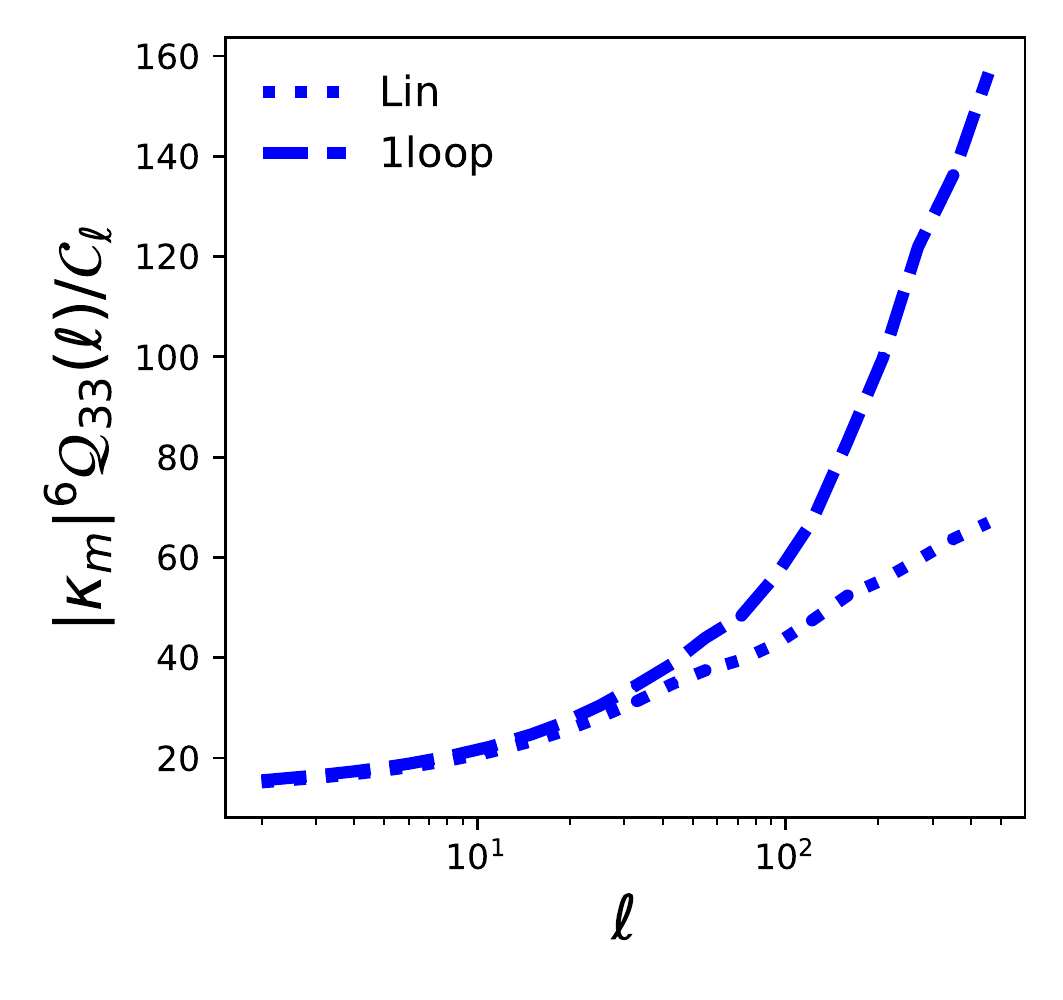}
  \end{minipage}
  \caption{Same as Fig-\ref{fig:3D_response} but for $z_s=2.0$}.
  \label{fig:3D_response_zs2}
  \end{center}
\end{figure}
%%%%%%%%%%%%%%%%%%%%%%%%%%%%%%%%%%
%
%
The {\em local} shear power spectrum denoted as
${\cal C}^{\gamma\gamma}_{\ell}(r_1,r_2 \big | \kappa_{L1}, \kappa_{L2})$
is given by:
\bes
\ben
    && {\cal C}^{\gamma\gamma}_{\ell}(r_1,r_2 \big | \kappa_{L1},\kappa_{L2})
    = {9\Omega^2_M H_0^4 \over 16 \pi^4 c^2}
    {(\ell+2)! \over (\ell-2)!}
    \int {dk \over k^2} V^{\gamma}_{\ell}(r_1,k | \delta_{L})
    V^{\gamma}_{\ell}(r_2,k | \delta_{L}) \label{eq:xla_gamma};\\
    && \kappa_{Li} =
    \delta_{Li} |\kappa_{mi}|;
    \quad \kappa_{mi} = -\int_0^{r_{si}} dr D_+(r)\, W_i(r).
    \een
    \ees
    Here, $\kappa_{mi}$, i.e. the minimum value of $\kappa_{Li}$ depends
    on the radial distance $r_i$ both through the maximum of the integration
    as well as through the weight $\rtrv{W_i}$. Notice that the global 3D spectrum
    ${\cal C}^{\gamma\gamma}_{\ell}(r_1,r_2)$ is recovered for $\delta_{L}=0$
    in which limit $\kappa_{Li}=0$. 
    The 
    spectroscopic surveys can measure the radial distances with higher accuracy
    but typically for fewer objects. In comparision photometric surveys
    target a higher number of galaxies, but in general with larger
    redshift uncertainities. The formalism here is suitable for photometric
    surveys. The quantities $V^{\gamma}$ in Eq.(\ref{eq:xla_gamma}) is expressed
    using a new function $U_{\ell}$
    \bes
    \ben
    && V^{\gamma}_{\ell}(\eta_i,k| \delta_L) \equiv \int \,dz_p\, dz^{\prime}
    n(z_p)\, p(z^{\prime} | z_p) W_i U_{\ell}(r[z_p],k);\\
    && U_{\ell}(r[z],k |\delta_L) \equiv
    {\int_0^r {dr^{\prime} a^\rtrv{-1}(r^{\prime})} {(r-r^{\prime}) \over r r^{\prime}}
      j_{\ell}(kr^{\prime}) P_{}^{1/2}(k; r^{\prime} | \delta_L)}.
\een
\ees
Here, $j_{\ell}$ is the spherical Bessel Function of order $\ell$.
The cross-power spectrum involving two different redshifts
(equivalently, two different radial distances) is often
factorized using the corresponding geometric mean, i.e.,
$P_{\delta}(k,r_1,r_2) = [P_{\delta}(k,r_1)P_{\delta}(k,r_2)]^{1/2}$. This approximation
reduces a higher-dimensional integral to a product of lower-dimensional
integrals. The accuracy of this ansatz was scrutinized in
\citep{Kitching_Heavens} in the context of weak lensing and
was found to be at the level of $10\%$ for scales $k>5 h^{-1}{\rm Mpc}$.
However notice that, using Zel\'dovich Approximation  ref.\citep{unequal2}, it was shown
that higher accuracy can be achieved. Indeed, we have a generalised
the factorization scheme by adopting it for local power spectrum, i.e.,
$P_{\delta}(k,r_1,r_2|\delta_L) = [P_{\delta}(k,r_1| \delta_L)P_{\delta}(k,r_2| \delta_L)]^{1/2}$.
Next, by Taylor expanding  ${P}^{1/2}_{\delta}(k,t|\delta_L)$: 

\ben
&& {P}^{1/2}_{}(k,t|\delta_L)= P^{1/2}_{}(k,t)\sum_{n=\rtrv{0}}^{\infty}
    {\delta_L^n \over n!} T_n(k,t).
 \label{eq:afunda2} 
\een
    The coefficients
    $T_n$ can be expressed in terms $R_n$ using Eq.(\ref{eq:afunda2}).
    \bes
    \ben
     && T_1 =  {1\over 2} R_1; \label{eq:T1}\\
     && T_2 =  -{1\over 2} R_1^2 + R_2; \\
     && T_3 =   {3\over 8} R_1^3 - {3\over 4} R_1 R_2 + {1\over 2} R_3. \label{eq:T3}
    \een
    \ees
    It is expected that the radius of convergence for the 
    Taylor expansion of $T^{1/2}$ will be smaller than the original 
    Taylor expansion of $T$.
    We will also Taylor expand the
    functions ${U}^{}_{\ell}$ and ${V}^{}_{\ell}$:
\ben
&& {U}^{}_{\ell}(k,t|\delta_L)=\sum_{n=\rtrv{0}}^{\infty}
    {\delta_L^n \over n!} U^{(n)}_\ell(k,t); \quad 
    {V}^{}_{\ell}(k,t|\delta_L)=
    \sum_{n=\rtrv{0}}^{\infty}
    {\delta_L^n \over n!}{V}^{(n)}_{\ell}(k,t).
    \een
    This will allow use to express the coefficients $U_\ell$ 
    in terms of the response functions $T_\alpha$, and subsequently
    $R_\alpha$.
    \ben
    U^{(n)}_\ell(k,t) ={1 \over |\kappa_m|^n}
    {\int_0^r {dr^{\prime} a^\rtrv{-1}(r^{\prime})} {(r-r^{\prime}) \over r r^{\prime}}
      j_{\ell}(kr^{\prime}) T_{n}(k,t)\, P_{}^{1/2}(k; r^{\prime})}.
    \een
    Using the Limber approximation
    $\lim_{\ell\rightarrow\infty}j_{\ell}(x)
    = \sqrt{\pi\over 2 (\ell+1/2)}\delta_D({\ell+1/2}-x)$
    we can simplify this
    to the following form \citep{Ilbert}:
    \ben
    U^{(n)}_{\ell}(r,k) = {1 \over |\kappa_m|^n}{r-{\cal L}(k) \over k a({\cal L}(k)) r {\cal L}(k)}
    \sqrt{\pi \over 2 (\ell+1/2)}T^{n}(k,{\cal L}(k))
    P_{}^{1/2}(k,{\cal L}(k));
    \een
    where we have used the following shorthand
    notation: ${\cal L}(k) = {(\ell + 1/2)/k}$.
    %\ben
    %G^{n}_\ell(k,t) = \int \,dz_p\, %dz^{\prime}
    %n(z_p)\, p(z^{\prime} | z_p) W_i %U^{n}_{\ell}(r[z_p],k). 
    %\een
    Taylor expanding the power spectrum in a bivariate series we define
    the response functions for the 3D power spectrum:
\ben
&& {\cal C}^{\gamma\gamma}_{\ell}(r_1,r_2\big | \kappa_{L1}, \kappa_{L2}) =
\sum_{a,b=0}^{\infty}{1\over a!}{1\over b!}\;
    {\cal Q}_{ab}(r_1,r_2, \ell) \;
    {\kappa^a_{L1}}{\kappa^b_{L2}}\;
    {\cal C}^{\gamma\gamma}_{\cal \ell}(r_1,r_2).
\label{eq:cls_response}
\een

The 3D response functions ${\cal Q}_{ab}$ of order
$ab$ is a function of
two source redshifts $r_1$ and $r_2$ and are given by:
\ben
    {\cal Q}_{ab}(r_1,r_2, \ell) = 
    {9\Omega^2_M H_0^4 \over 16 \pi^4 c^2}
    {(\ell+2)! \over (\ell-2)!}
     \int {dk \over k^2} G^{a}_{\ell}(r_1,k)
     G^{b}_{\ell}(r_2,k).
     \label{3D_response}
    \een
    For the radial distribution of galaxies denoted by $n(z)$ typically 
    the following form is considered:
    \bes
    \ben
    && n(z) = (z/z_e)^2 \exp[-(z/z_e)^{3/2}]; \quad z_e = 0.9/\sqrt{2}.
    \een
    The photometric smoothing is represented by the following
    Gaussian photometric distribution.
    \ben
    && p(z|z_p) = {1\over 2\pi \sigma_z(z_p)}
    \exp \Big [-{1\over 2 \sigma_{z_p}} (z-c_{cal}z_p +z_{\rm bias})^2 \Big ]; \nn
    && c_{cal} = 1.0; \quad z_{\rm bias}=0.0; \quad
    \sigma_{z_p} = A(1+z_P); \quad A=0.5.
    \een
    \ees
     For the results shown for
     ${\cal Q}_{ab}$ in Figure-\ref{fig:3D_response}
and Figure-\ref{fig:3D_response_zs2}, we assume a single source redshift ($z_s=1$ or $2$) instead of the source distribution. We also neglect the photo-z error and replaced $p(z|z_p)$ it with a delta function.
%
%
%
%
%
%%%%%%%%%%%%%%%%%%%%%%%%%%%%%%%%%%%
\section{$k$-cut Response Functions}
\label{sec:kcut}
%%%%%%%%%%%%%%%%%%%%%%%%%%%%%%%%%%%

As is well known, the cosmic shear statistics is very sensitive
to small scale power which depends on
poorly understood nonlinear physics as well as baryonic feedback.
Many techniques have been
developed from brute force N-body simulation to model
small scale behaviour with subsequent marginalisation over small scale
power spectra to develop emulator based approach
that can be combined with fast  Monte Carlo Markov Chain schemes.
 However, each of this techniques
are either too expensive or lacks sufficient accuracy required for
stage-IV experiments.

A solution to this problem was first proposed in \cite{photoz} (see also 
\citep{kcut})
that is geometric in nature and cuts out the weak 
lensing spectrum’s sensitivity to small scale structure
in a tunable power. We will refer
to the power spectra computed in this manner as k-cut
power spectra. This method relies on a nulling scheme
that is achieved by applying a similarity
transform to the weak
lensing spectra following \citep{BNT}.
The key aspect of this transformation is that it organises
the lensing information in the lens plane instead of the
source plane. Next, taking advantage of the fact that
each bin constructed in this manner corresponds to a particular lens
redshift range, so taking an angular scale cut thus also
removes sensitivity to large-k (small scales) in a uniform manner. 
In this section we generalise the idea idea of 
k-cut spectra to k-cut response functions \citep{kcut, BNT}.

If we consider a set of discrete source
planes at radial distances $r_i$,
\ben
&&  {\tilde W}_{\alpha}(r) = \sum_i p^i_\alpha W_i
= {3 \Omega_M H_0^2\over 2 c^2} a^{-1}(r)\sum_{i;\; r_i > r} p^{i}_\alpha  { d_A(r_i-r)d_A(r) \over d_A(r_i)},
\een
where $\{p_i\}$ are a set of weights associated with source planes.
The key step in implementing the 
nulling scheme introduced in \citep{BNT} is to select weights in
such a manner that is the weighted convergence $\tilde \kappa_\alpha$
is only sensitive to lenses in a specific radial distance.
\bes
\ben
\tilde \kappa_\alpha = \sum_i p^i_\alpha \kappa_i.
\een
In the harmonic domain:
\ben
\tilde \kappa_{\alpha, \ell m} = \sum_i p^i_\alpha \kappa_{i,\ell m};
\een
\ees
We will refer to above transformation as the Bernardeau-Nishimichi-Taruya (BNT) transformation,
and $\tilde \kappa_a$ as BNT transformed convergence.
It can be shown that the BNT weighted power spectra denoted as
$\tilde{\cal C}^{\alpha\beta}_{\ell}
\equiv \langle\kappa_{\ell m}^\alpha\kappa_{\ell m}^{\beta*} \rangle$ is related to the
ordinary tomographic spectra ${\cal C}^{ij}_{\ell} \equiv \langle\kappa_{\ell m}^i\kappa_{\ell m}^{*j} \rangle $ through the following
similarity (BNT) transformation.
\ben
\tilde{\cal C}^{\alpha\beta}_{\ell} =  \sum_{i,j} p^i_\alpha p^j_\beta {\cal C}^{ij}_{\ell}.
\een
In a more compact matrix notation we can express the
similarity transform as:
\ben
\tilde{\bf C}_{\ell} = {\bf M} {\bf C}_{\ell} {\bf M}^T.
\een
Construction of the transformation matrix ${\bf M}$ from the weights  $p^i_\alpha$,
which satisfies various constraints, is detailed in \citep{BNT,kcut}: 
\ben
&& \tilde{\cal C}^{\alpha\beta}_{\ell}(\tilde\kappa_{LX} ) =  \sum_{n=0}^{\infty}{1\over n!}\;
    \tilde{\cal Q}^{\alpha\beta}_n(\ell) \; {\tilde\kappa^n_{LX}} \tilde{\cal C}^{\alpha\beta}_{\cal \ell}.
\label{eq:cls_cross_response}
\een
This is the BNT equivalent of Eq.(\ref{eq:cross_spectra_define}).
We can now define the position dependent
BNT transformed tomographic spectra $\tilde{\cal C}^{\alpha\beta}_{\ell}(\tilde\kappa_{LX} )$
that depends on $\tilde\kappa_{LX}\equiv [\tilde\kappa_{L\alpha}\tilde\kappa_{L\beta}]^{1/2}$.
Going through the algebra we find the equivalent of Eq.(\ref{eq:def_QN}):
\ben
{\tilde Q}^{\alpha\beta}_n(\ell) = {1\over {\cal C}^{\alpha\beta}_{\ell}} {1  \over |{\tilde\kappa}_{LX}|^n}
\int_0^{r_{s}} d\tilde r {{W}^\alpha(r){W}^\beta(r) \over  {d^2_A}(r)}
R_n\left [{\ell \over d_A(r)},r\right ][D_+(r)]^n
P_{\delta}\left ({\ell\over d_A(r)},r \right ).
\label{eq:def_QNab}  % {eq:def_QN} 
\een
The corresponding expression for tomographic binning is given in Eq.(\ref{eq:def_QN}).
Notice the normalisation of the k-cut response function ${\tilde Q}^{\alpha\beta}_{\ell}$ is
different from that of the ordinary response functions ${Q}^{\alpha\beta}_{\ell}$.
The following definitions were used to express $\tilde\kappa_{LX}$:
\bes
\ben
&& \tilde\kappa_{\alpha L} =  \delta_L |\kappa_{\alpha m}|; \quad\quad
\tilde\kappa_{\alpha m} = \sum_i p^i_\alpha \kappa_{im}(r) = -\int_0^{r_s} dr\, D_+(r)\,\sum_i p_i  w_i(r).
\een
\ees
Applying a suitable $\ell$ cut-off in Eq.(\ref{eq:def_QN}) we can systematically remove the
high-$k$ modes.
`
%%%%%%%%%%%%%%%%%%%%%%%%%%%%%%%%%%%
\section{Results and Discussions}
\label{sec:results}
%%%%%%%%%%%%%%%%%%%%%%%%%%%%%%%%%%%%
%
In Figure-\ref{fig:QN_z1} we show the response functions
 $Q_N$ defined in Eq.(\ref{eq:def_QN}) are shown. From
  left-to right panels depict $N=1,2$ and $N=3$.
  The source redshift is at $z_s=1$.
        Various line-styles correspond to different analytical
    models, linear, one-loop and halo model as indicated.
    In Figure-\ref{fig:QN_z2} we present the
    corresponding results for
    the source redshift $z_s=2.0$.  
    The predictions from one-loop SPT for 
    lower order response functions
    show relatively higher level of agreement with the HM models

Typically, for the intermediate range of $\ell$ values most
models show an increasing trend. While the HM actually shows
a declining trend the predictions based on 1-loop saturates
at a rather high value. The predictions based on linear theory
are relatively more stable. While each of these predictions
need to be checked against simulations, the response function
technique based on power spectrum that can probe 
squeezed bispectrum are more easy to implement 
compared to the detailed modelling of the bispectrum.
    
    In Figure-\ref{fig:R1} we show the corresponding response
    functions for the underlying matter distribution $z=1.$
    The response functions $R_N$ defined in
    Eq.(\ref{eq:R1})-Eq.(\ref{eq:R4}) are shown. From
    left to right panels depict $N=1,2$ and $N=3$. For the
    response function $R_1$ various models agree with each other
    for $k< h\,Mpc^{-1}$. The disagreement among them is more
    pronounced at lower $z$ and higher $N$.
    
In Figure-\ref{fig:QN_corr} we show the response functions for the
    correlation function. The source redshift is fixed at $z_s=1.$
  The response functions $\Sigma_N$ defined in  Eq.(\ref{eq:response})
  are shown for the source redshift $z_s=1.0$. From
    left to right panels depict $N=1,2$ and $N=3$. 
    Various line styles correspond to different analytical
    models, linear, one-loop and halo model as indicated.
As expected for large separation angle $\theta$ all models 
show similar trends, but they differ in the small separation
regime.

In Figure-\ref{fig:TN_z2} we have plotted the
$T_N$ parameters defined in Eq.(\ref{eq:T1}) - \ Eq.(\ref{eq:T3}).
These coefficients can be obtained by Taylor expanding
square roots of the ratio of local and global power spectrum
$P^{1/2}(k,r | \delta_L)/P^{1/2}(k,r) $
and are related to the coefficients $R_N$ and are
functions of the wave number $k$. The trends in $T_N$ with 
$z$ and $N$ is dictated similar trends in $R_N$.

The Figure-\ref{fig:3D_response} shows 3D
response functions. The 3D response function is defined in Eq.(\ref{eq:cls_response}). In 3D the response function
depends on two different source redshiftd $z_{s1}=1$ and $z_{s2}=2$. From left to right panels
depict $Q_{11}$, $Q_{22}$ and $Q_{33}$. We show  these results for linear theory and 1-loop SPT. In agreement with their projected counterparts one-loop corrections
show departure at increasing lower $\ell$. 
    
%%%%%%%%%%%%%%%%%%%%%%%%%%%%%%%%%%%%%%%%
\section{Conclusion and Future Prospects}
\label{sec:conclu}
%%%%%%%%%%%%%%%%%%%%%%%%%%%%%%%%%%%%%%%%

Several authors in recent years have used SU
formalism in the context of galaxy clustering studies \rtrv{(e.g. \citep{chiang, Wagner,terasawa}}).
In this paper we have introduced
the response functions approach for
analysing the higher-order statistics
of weak lensing convergence maps.
We have also extended the real space based correlation function
results \citep{BOSS} developed for galaxy surveys
for the case of weak lensing surveys.
The response functions for the correlation
functions presented here can be generalised
to $3\times 2$ correlation functions typically
used to analyse the data weak lensing surveys.
For a different approach to
response function see \citep{response}. %BERNARDEAU_Response}.

We have explored the response functions for
weak lensing power spectrum.
However, the formalism discussed
here can be generalised  
for bispectrum and other higher-order statistics.
Separate Universe N-body simulations for dark matter clustering
are currently available, but separate universe weak lensing convergence
or shear maps from such simulations are currently
unavailable. We hope our study will motivate
development of such simulations.
The validity range of various approximations
used in our derivation can then be tested when such simulations become available.
                                   
The forward modelling studies based on
power spectrum have gained popularity
in recent years.
These studies can be extended
to include the information regarding
non-Gaussianity using the response functions
introduced here without much
additional computational
overhead.

To compute the signal-to-noise associated
with the response functions we have studied
here, the covariance matrices for these statistics
is needed, which will be presented in a separate
publication.

The preferential alignments of halos due
to tidal interactions
is responsible for what is also known
as intrinsic alignment (IA)
and is considered to be a systematics for weak lensing
surveys see \citep{Kids_Joachimi} for
KiDS and \citep{Secco} for DES.
It is believed that for analysing
the future surveys such as Euclid and LSST
it will be vital to understand
IA in a lot more detail.
Many authors on the other hand
have gone a step forward and underlined
the usefulness of IA as a cosmological probe.
Most statistical modelings of IA is devoted
to halo model based approaches. In \citep{EFT}
an effective field theory (EFT) based
approach was developed for modelling
of power spectrum and in \citep{intrinsic}
the authors have focused on bispectrum
induced by IA. A response function based
approach that only relies on modelling of
power spectrum and its derivatives
will be presented elsewhere.

The theoretical framework developed here will also be
useful beyond weak lensing studies in other areas
of cosmology, e.g., in the context of
Lyman-$\alpha$ \citep{Lyman}
absorption studies,
21cm studies \citep{21cm} and studies of
CMB secondaries\citep{ksz}.

%%%%%%%%%%%%%%%%%%%%%%%%%%
\section*{Acknowledgment}
%%%%%%%%%%%%%%%%%%%%%%%%%%
DM was supported by a grant from the
Leverhulme Trust at MSSL where
this work was initiated. It is a pleasure for
DM to acknowledge an Advanced Research Fellowship
at Imperial Centre for Inference and Cosmology (ICIC) where this work was completed.
We would like to thank Alan Heavens for careful
reading of the draft and many constructive comments.
This work was supported by JSPS KAKENHI Grant Numbers JP22H00130 and JP20H05855 (RT).
%
%%%%\bibliography{sep1}

\appendix
%%%%%%%%%%%%%%%%%%%%%%%%%%%
\section{Perturbative Results}
\label{sec:perturbative}
%%%%%%%%%%%%%%%%%%%%%%%%%%%%%% 
Following \citep{IB}, the expression for the exact 2D expression is given by:
\bes\ben
&& S_3(\ell) = N_2\left [ {24\over 7} -{1\over 2}(n+1) \right ] \\
&& N_2 = {\int_0^{r_s}} dr {\omega^3(r) \over d^{4+2n}_A(r)} \Big /
\left ( {\int_0^{r_s}} dr {\omega(r) \over d^{4+2n}_A(r)} \right )^2
\een\ees
and also the doubly squeezed trispectrum is given by:
\bes\ben
&& S_3(\ell) = N_3\left [ {1473\over 79} -{195\over 14}(n+2)
  + {3\over 4}(n+2)^2 \right ] \\
&& N_3 = {\int_0^{r_s}} dr {\omega^4(r) \over d^{6+3n}_A(r)} \Big /
\left ( {\int_0^{r_s}} dr {\omega^2(r) \over d^{6+3n}_A(r)} \right )^2
\een\ees

\end{document}